\documentclass[11pt]{article}
\usepackage{epsf}
\usepackage{axodraw}

\newcommand{ \centeron }[2]{{\setbox0=\hbox{#1}\setbox1=\hbox{#2}\ifdim
                             \wd1>\wd0\kern.5\wd1\kern-.5\wd0\fi \copy0
                             \kern-.5\wd0\kern-.5\wd1\copy1\ifdim\wd0>\wd1
                             \kern.5\wd0\kern-.5\wd1\fi}}
\newcommand{ \ltap }{\>\centeron{\raise.35ex\hbox{$<$}}
                     {\lower.65ex\hbox{$\sim$}}\>}
\newcommand{ \gtap }{\>\centeron{\raise.35ex\hbox{$>$}}
                     {\lower.65ex\hbox{$\sim$}}\>}
\newcommand{ \gsim }{\mathrel{\gtap}}
\newcommand{ \lsim }{\mathrel{\ltap}}

\newcommand{ \slashchar }[1]{\setbox0=\hbox{$#1$}   
   \dimen0=\wd0                                     
   \setbox1=\hbox{/} \dimen1=\wd1                   
   \ifdim\dimen0>\dimen1                            
      \rlap{\hbox to \dimen0{\hfil/\hfil}}          
      #1                                            
   \else                                            
      \rlap{\hbox to \dimen1{\hfil$#1$\hfil}}       
      /                                             
   \fi}                                             %

\newcommand{ \ratio    }[1]{M_{#1}/\alpha_{#1}}

\newcommand{ \tr       }{\mathop{\mathrm{tr}}}

\newcommand{ \ra       }{\rightarrow}
\newcommand{ \re       }{\mathop{\mathrm{Re}}}

\newcommand{ \textfrac }[2]{ {\textstyle\frac{#1}{#2}} }

\def\singleandabitspaced{\baselineskip=\normalbaselineskip\multiply
    \baselineskip by 104\divide\baselineskip by 100}
\def\singlespaced{\baselineskip=\normalbaselineskip}

\newcommand{\Journal}[4]{{#1}\ \textbf{#2}, #3 (#4)}  
\newcommand{ \NPB    }[3]{\Journal{Nucl. Phys.}{B#1}{#2}{#3}}
\newcommand{ \PLB    }[3]{\Journal{Phys. Lett. B}{#1}{#2}{#3}}

\newcommand{ \PRD    }[3]{\Journal{Phys. Rev.}{D#1}{#2}{#3}}

\newcommand{ \PRL    }[3]{\Journal{Phys. Rev. Lett.}{#1}{#2}{#3}}
\newcommand{ \PREP   }[3]{\Journal{Phys. Rept.}{#1}{#2}{#3}}
\newcommand{ \ZPC    }[3]{\Journal{Z. Phys. C}{#1}{#2}{#3}}

\newcommand{ \PTP    }[3]{\Journal{Prog. Theor. Phys.}{#1}{#2}{#3}}

\newcommand{ \PPNP   }[3]{\Journal{Prog. Part. Nucl. Phys.}{#1}{#2}{#3}}

\newcommand{ \xxx    }[1]{\texttt{[#1]}}

\begin{document}

\singlespaced

\begin{titlepage}

\begin{flushright}
hep-ph/9803259 \\
UM-TH-98-05 \\
March 1998
\end{flushright}

\vspace{0.8cm}

\begin{center}
\mbox{\Large \textbf{Disrupting the one-loop renormalization group 
                 invariant $M/\alpha$}} \\
\vspace*{0.3cm}
\mbox{\Large \textbf{in supersymmetry}} \\

\vspace*{1.6cm}
{\large Graham~D.~Kribs} \\
\vspace*{0.5cm}
{\it Randall Physics Laboratory, University of Michigan} \\
\vspace*{0.00cm}
{\it Ann Arbor, MI~~48109-1120} \\
\vspace*{0.5cm}
{\tt kribs@feynman.physics.lsa.umich.edu}
\vspace*{0.6cm}

\begin{abstract}
\indent

\singleandabitspaced

It is well known that in low energy supersymmetry the ratio 
of the gaugino mass to the gauge coupling squared, $M/\alpha$, 
is renormalization group invariant to one-loop.
We present a systematic analysis of the corrections
to this ratio, including standard two-loop corrections 
from gauge and Yukawa couplings, corrections 
due to an additional $U(1)'$ gaugino, threshold corrections, 
superoblique corrections, corrections due to extra matter, 
GUT and Planck scale corrections, and ``corrections'' from messenger 
sectors with supersymmetry breaking communicated via gauge-mediation.
We show that many of these effects induce corrections at the level 
of a few to tens of percent, but some could give much larger corrections, 
drastically disrupting the renormalization group extrapolation 
of the ratio to higher scales.  
Our analysis is essentially model-independent, and therefore
can be used to determine the ambiguities in extrapolating the
ratio in any given model between the weak scale and higher scales.

\end{abstract}

\end{center}
\end{titlepage}

\newpage
\setcounter{page}{2}
\renewcommand{\thefootnote}{\arabic{footnote}}
\setcounter{footnote}{0}
\singleandabitspaced

\section{Introduction}
\label{introduction-sec}
\indent

It is well known that any hope of stabilizing a hierarchy of
scales requires supersymmetry~\cite{reviews}.  The disparity between 
the weak scale and the grand unified theory (GUT) scale
or the Planck scale (hereafter lumped together
as the ``high scale'') is bridged in a softly broken 
supersymmetric standard model by rendering scalar mass
renormalization to be at most logarithmically divergent.
The canonical approach to connecting these disparate scales
is through the renormalization group (RG), and there has been
enormous effort in calculating and evaluating the RG 
equations analytically and numerically,
e.g.\ Refs.~\cite{RR, JMY, BBO1, LP1, CPW,
Pokorskibottomup, CPR, BBO2, KKRW, LNZ, CPP}.
Most of this effort has been directed towards either
the consequences of a given high scale theory 
(be it a GUT, string theory, or other high scale proposal) 
on weak scale phenomenology, or the consequences of present 
(or proposed) weak scale measurements on high scale physics.

A general, softly broken, weak scale supersymmetric
model has a large number of additional parameters beyond those
of the standard model (SM)~\cite{reviews}.
Most of the parameters are interconnected through the RG 
equations, and therefore successful extrapolation generally
requires a simplifying framework (or an organizing principle) 
to be imposed either at the weak scale or the high scale.
Even with the varied simplified frameworks that have been 
traditionally used, such as supergravity-motivated models, 
gauge-mediated models, and so on, there are a sufficient number 
of parameters that one usually resorts to numerical sampling.
While this approach usually gives the correct general trends,
it could also easily give misleading results if the physics
of the model depends very sensitively on a sampled parameter.

Disentangling the dependencies of any given supersymmetric parameter 
on other parameters has proved to be formidable task.
The first step, extracting weak scale observables, is likely going 
to be challenging both experimentally (discovery, and then mass and
coupling measurements) and theoretically (mixings between gauginos, 
one-loop corrections, etc.).  Hence, the subsequent extrapolation 
of the full theory to higher scales will be unreliable unless
quantities can be found that are not highly interdependent on 
other parameters of the theory.  One such quantity is the ratio
of the gaugino mass to the gauge coupling squared, $\ratio{a}$, 
where $a=1,2,3$ corresponds to $U(1)_Y, SU(2)_L, SU(3)_c$.  
In principle, one can extract $\ratio{a}$ by ``merely'' measuring 
the gaugino mass, given that the standard model gauge couplings
are by now well measured.  Remarkably,
the one-loop RG equation for $\ratio{a}$ vanishes~\cite{IKKT},
which suggests that a measurement at the weak scale would
determine the value at the high scale, up to small two-loop
corrections (albeit scaled by 
$\sim \ln Q_{\mathrm{high}}/Q_{\mathrm{weak}}$).

The usefulness of the ratio $\ratio{a}$ at higher scales is,
ultimately, model-dependent.  However, there are strong 
motivations to think that the pattern of the gauge couplings
and the gaugino masses at the high scale will determine a great 
deal about the high scale theory.  In GUTs, one ordinarily 
expects that the gauge couplings and the gaugino masses will ``unify'',
or take the same value, after being embedded in the GUT 
group~\cite{Mohapatra}.  Even in string theory there
are reasons to suggest the gaugino masses could be
unified in the simplest scenarios~\cite{string-gaugino-simple}.  
If we could reliably calculate $\ratio{a}$ at the high scale, 
no doubt one could match string phenomenology 
to such quantities, strengthening the predictions 
of successful models (and eliminating classes of unsuccessful models).
In addition, given a motivation for thinking $g_{\mathrm{GUT}}$ or
$g_{\mathrm{string}}$ should take a certain value, it would
also be trivial to extract the gaugino masses by themselves 
from the ratios $\ratio{a}$.  

The difficulty of unambiguously extrapolating the individual quantities
$g_a$ and $M_a$, or even other weak scale parameters such as
squark and slepton soft masses, is mainly due to the one-loop
RG evolution and one-loop dependence on threshold corrections.
In particular, extrapolating up $13$ orders of magnitude in
scale to the apparent unification scale $M_{\mathrm{unif}} 
\approx 2 \times 10^{16}$~GeV is notoriously complicated 
by weak scale supersymmetry threshold corrections, GUT scale 
threshold corrections, extra matter at intermediate scales, 
Planck scale corrections, etc.  
Using the ratio $\ratio{a}$ mitigates many of these 
issues due to the one-loop RG invariance, as we will see.

Our goal is ultimately to determine how well 
$\ratio{a}$ can be known at higher scales, 
and thus we will be primarily concerned with 
running \emph{up} to the high scale (or messenger scale), rather 
than the more traditional approach of running \emph{down} from 
a boundary condition imposed by some supersymmetry 
breaking ansatz.  Our results can, of course, be easily inverted 
should there be a motivation to do so.  
Bottom-up approaches to supersymmetry have been considered
before (see e.g.\ \cite{Pokorskibottomup, Zichichibottomup, 
CWbottomup, gordytalks}), generally with the philosophy that
weak scale phenomenology ought to be extrapolated to the
high scale, leaving the high scale matching to GUTs or string theory.
Our approach is similar, in that we start from the weak scale,
but differs from some of the previous analyses by extrapolating
only the ratios $\ratio{a}$, and yet also considering various
high scale effects.  Throughout most of this work
we illustrate the results of extrapolation from near the weak
scale $\sim 1$~TeV up to the apparent unification scale 
$\sim 10^{16}$~GeV; should the soft mass generation scale be 
much lower as in gauge-mediated supersymmetry breaking models, 
the size of the corrections will be smaller, but can still be 
easily extracted from most of the graphs we present.  Similarly, 
effects in high scale theories (with gravity-mediated supersymmetry 
breaking) can give corrections to the ordinary expectations for 
$\ratio{a}$ at the weak scale.  By quantifying the various effects 
that could disrupt the one-loop expectations for $\ratio{a}$, 
we are simultaneously showing that a naive interpretation of 
the gaugino mass ratios at the weak scale could easily give misleading 
conclusions, but a systematic evaluation of the effects discussed 
here could lead to intriguing signals of physics at much higher
scales.

The most obvious class of corrections to the one-loop 
invariant $\ratio{a}$ are the two-loop 
corrections~\cite{YamadaGaugino, MVGaugino, YamadaRatio} 
(in the supersymmetric $\overline{\mathrm{DR}}$ renormalization 
scheme~\cite{DRED}).
The two-loop corrections ought to be small,
if perturbation theory is valid\footnote{Which we assume
throughout this paper.}.  However, there are reasons to
suggest that the size of two-loop corrections
could be larger than one might naively guess.  The two-loop 
coefficients are not of $\mathcal{O}(1)$, but in fact $\mathcal{O}(10)$ 
(in the minimal supersymmetric standard model (MSSM) 
the two-loop coefficients lie in the range 
$9/5 \le B_{ab}^{(2)} \le 25$).  Also, two-loop terms proportional
to the Yukawa couplings are present, and of course the
top Yukawa is large at the weak scale.
Finally, extrapolation to the
high scale involves running over $13$ orders of magnitude in scale, 
and therefore the (resummed) logarithm from RG evolution is 
comparatively large.  (Analogously, if the soft mass generation
scale is much lower, then the effect of RG evolution 
is not nearly as large.)

If two-loop corrections become important, then three-loop corrections
may also be important, if for no other reason than to check
that perturbation theory is indeed valid.
However, a complete three-loop analysis would entail using 
three-loop gauge and gaugino mass $\beta$-functions,
at least two-loop Yukawa couplings and scalar trilinear
couplings, and two-loop thresholds.  The three-loop
gauge coupling $\beta$-functions for a general MSSM were 
presented in Ref.~\cite{three-loop-gauge}, and the three-loop
gaugino $\beta$-functions were recently presented in
Ref.~\cite{three-loop-gaugino}.  Two-loop $\beta$-functions
for the Yukawa couplings and scalar trilinear couplings
are by now well known~\cite{MVFull, YamadaFull}.
To the best of our knowledge two-loop threshold corrections 
(in a nondecoupling scheme) have not been calculated, 
although one could implement
a decoupling scheme whereby the two-loop coefficients
are changed as each threshold is crossed; this
procedure is an approximation that should be straightforward
to calculate from generalized two-loop $\beta$-functions,
although we decline to present it here.  An additional well-known 
complexity in going to three-loops is that the $\beta$-functions 
are no longer scheme independent~\cite{scheme-dependence}.
In any case, we will not attempt to calculate corrections 
beyond two-loop, instead relying on others' 
calculations~\cite{three-loop-gauge, KoldaMR}
of three-loop corrections to guide us in those cases 
where we suspect perturbation theory may be in trouble.

The central reason why three-loop corrections should be unnecessary
(unless a coupling gets large) is that there is no dependence on any
additional parameters beyond those needed for the two-loop corrections.  
The intriguing pattern of the supersymmetric parameter interdependence
is given in Table~\ref{depend-table}, where we show the
dependencies arising at one-, two-, and three-loop order
in the RG equations.
\begin{table}
\renewcommand{\baselinestretch}{1.2}\small\normalsize
\begin{center}
\begin{tabular}{r|ccc}
     & One-loop & Two-loop & Three-loop \\ \hline
$g_i$  & $g_i$ & $g_a, Y_x$ & $g_a, Y_x$ \\
$Y_w$  & $Y_x, g_a$ & $Y_x, g_a$ & $Y_x, g_a$ \\
$M_i$  & $M_i, g_i$ & $M_a, g_a, Y_x, A_x$ & $M_a, g_a, Y_x, A_x$ \\
$A_w$  & $A_x, g_a, Y_x, M_a$ & $A_x, g_a, Y_x, M_a$ & $A_x, g_a, Y_x, M_a$ \\
\end{tabular}
\end{center}
\renewcommand{\baselinestretch}{1.0}\small\normalsize
\caption{Dynamical parameter dependence in supersymmetry for 
one-, two-, and three-loop renormalization group equations.
In the MSSM the labels $i, a = 1, 2, 3$ specify gauge couplings
and $w, x = u, d, e$ specify Yukawa couplings,
with $i$ and $w$ fixed while $a$ and $x$ are summed over.}
\label{depend-table}
\end{table}
The results to three-loop have been explicitly 
calculated for the gauge couplings $g$, Yukawa couplings $Y$,
and gaugino masses $M$~\cite{three-loop-gauge, three-loop-gaugino},
and we suspect the three-loop result for the scalar trilinear 
couplings $A$ holds based on two-loop results~\cite{MVFull, YamadaFull}.

The parameter dependencies in Table~\ref{depend-table} are
central to our analysis.  They imply, specifically, that
the set of parameters $(g, M, Y, A)$ are sufficient to compute RG 
evolution to three-loop (and probably to all orders), and that
we need not be concerned with the squark or slepton soft masses 
or the bilinear couplings (such as $\mu$ or $B$) 
directly\footnote{The extraction of the scalar trilinear 
couplings at the weak scale, however, can presumably be obtained 
only by measuring the off-diagonal elements of the sfermion mass 
matrices.}.  This interdependence does, however, deserve a few 
further comments.

In the exact supersymmetric limit, only gauge couplings and 
superpotential parameters (including Yukawa couplings) remain.  
It is therefore not surprising that the gauge couplings 
and Yukawa couplings depend on each other, but not on any 
soft breaking parameter.  That the soft breaking parameters
$M$ and $A$ depend on only the four parameters $(g, M, Y, A)$
can ultimately be related to theoretical properties of the
superpotential.  There has been a significant renewed 
interest in exact $\beta$-functions in ($N=1$) supersymmetric 
theories.  Using a fully Wilsonian treatment, exact one-loop 
$\beta$-functions for the gauge couplings~\cite{exact-gauge} 
and gaugino masses~\cite{HisanoShifman, three-loop-gaugino} 
can be found by expressing the soft breaking masses as spurions and 
then exploiting the holomorphy of the Lagrangian.  Although these 
results are not directly useful for calculations with canonically 
normalized fields, they do allow the computation of higher order 
$\beta$-functions including some of the three-loop results
mentioned above.

The outline of the paper is as follows.  In Sec.~\ref{MSSM-sec}, 
we present two-loop effects in the MSSM\@.  We discuss the
two-loop RG equations including both the ``pure gaugino''
and Yukawa terms, and the dependencies on the sign of
the scalar trilinear coupling as well as the gaugino mass.
We also discuss weak scale threshold corrections and 
superoblique corrections to the ratio $\ratio{a}$.
In Sec.~\ref{beyond-sec} we discuss a series of effects
beyond the MSSM, including an extra $U(1)'$.  Consistently adding
an extra $U(1)'$ to the gauge structure of the MSSM 
requires the nontrivial cancellation of $U(1)'$
anomalies, and we relegate of discussion of this to 
Appendix~\ref{anomalies-app}.  We consider two models 
(with anomaly cancellation); a minimal $U(1)'$ model and
an $E_6$ model.  We also consider adding extra matter
to the theory, and its effect on the ratio $\ratio{a}$.
In Sec.~\ref{high-scale-sec} we discuss high scale corrections
to the ratio $\ratio{a}$ from GUT threshold corrections, 
effects due to breaking GUTs at other scales, Planck scale corrections, 
and supergravity effects. 
In Sec.~\ref{gauge-mediation-sec} we discuss the consequences
of extra matter with supersymmetry breaking masses that
induce ``corrections'' via gauge-mediation.  Finally, in
Sec.~\ref{conclusions-sec} we present our conclusions.
In Appendix~\ref{RGE-app}, we write the full two-loop RG
equations for the gauge couplings and gaugino masses
including all Yukawa couplings appropriate to the model, 
and then apply the results to the MSSM, MSSM $+ \> U(1)'$ model, 
and an $E_6$ model.

\section{Effects in the MSSM}
\label{MSSM-sec}
\indent

The first class of corrections to consider are those
arising from the ``standard'' two-loop terms in the
RG equation for $\ratio{a}$.  This will be done
by starting with certain restrictive assumptions
about the initial conditions and the two-loop terms
to gain insight into the corrections, and then removing 
the assumptions one by one.  Both weak scale threshold
corrections and superoblique corrections are considered here,
since they are generic in any weak scale supersymmetric
model.

\subsection{Two-loop effects in the MSSM}
\indent

\begin{figure}[!t]
\centering
\epsfxsize=4.0in
\hspace*{0in}
\epsffile{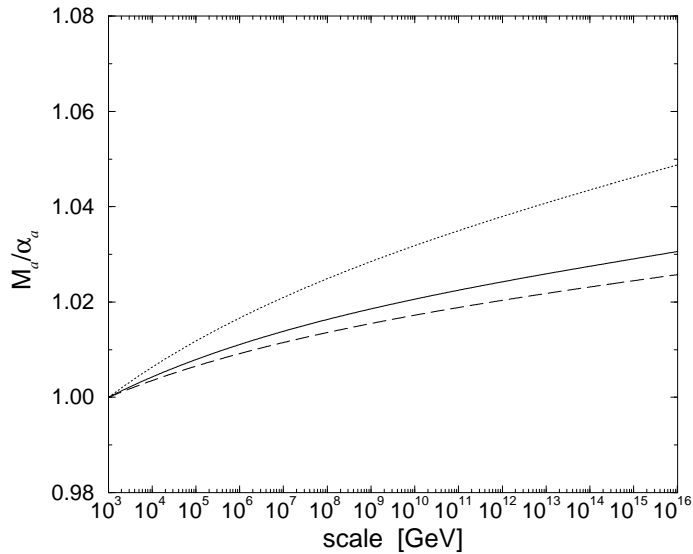}
\caption{Evolution of $\ratio{a}$ as a function of scale.
The solid, dotted, and dashed lines correspond to $a=1, 2, 3$
respectively.  Note that we have normalized the $y$-axis by dividing 
by the value $\ratio{1}$ at $1$ TeV.}
\label{case1-fig}
\end{figure}

To begin and start with a point of reference, consider
the RG evolution of $\ratio{a}$ using the two-loop RG equations 
in the MSSM\@.
The RG equation for the ratio $\ratio{a}$ is given in 
Eq.~(\ref{ratio-RGE-eq}), and throughout much of the following
discussion we use ``pure gaugino'' and ``Yukawa'' contributions 
to refer to the first and second set of terms in the square brackets 
of the RG equation.  First, take the unrealistic case where 
the Yukawa couplings are set to zero, which will allow us to 
analyze the pure gaugino contributions to the RG equations 
with the least number of complexities.
Furthermore, let's start by assuming that 
\begin{equation}
\ratio{1} \; = \; \ratio{2} \; = \; \ratio{3}
\label{init-cond-eq}
\end{equation}
holds at a scale $Q$ near the weak scale, 
but above all weak scale thresholds
(these will be discussed in Sec.~\ref{weak-thresholds-sec}). 
The evolution of the gauge couplings and gaugino masses 
is governed by six coupled RG equations
(see Appendix~\ref{RGE-app}), but this
is still analytically rather complicated.
In Fig.~\ref{case1-fig}, we show the evolution of the
ratio $\ratio{a}$ from $10^3 \ra 10^{16}$~GeV\@.
By choosing to start the running at $1$ TeV with Eq.~(\ref{init-cond-eq})
as the weak scale boundary condition, we are restricting
$M_1(1 \; \mathrm{TeV}) \lsim 190$ GeV, to prevent the gluino
mass $M_3(1 \; \mathrm{TeV})$ from exceeding $1$~TeV\@.  
For smaller values of $M_1$ (and correspondingly $M_2$ and $M_3$), 
the results are identical since there is only one scale 
in the problem (any one of the gaugino masses determines 
the scale).  Nevertheless, for definiteness we take
$M_1(1 \; \mathrm{TeV}) = 100$~GeV,
$M_2(1 \; \mathrm{TeV}) \approx 190$~GeV, and
$M_3(1 \; \mathrm{TeV}) \approx 527$~GeV in the following 
unless otherwise specified.  The latter values ($M_2$ and $M_3$)
are considered mildly approximate since their determination requires
the gauge couplings at $1$~TeV, and therefore subject to weak
scale threshold corrections.

There are a few important remarks to be made based on
this simple exercise.  First, the size of the pure gaugino
corrections to the one-loop invariant $\ratio{a}$
are less than $5$\% after scaling $13$ orders of magnitude.
In all cases the correction is a slight increase with increasing
scale (RG equation is positive), reflecting the usual choice 
of phases in the initial condition, Eq.~(\ref{init-cond-eq}).  The most
striking feature is that the size of the two-loop pure gaugino 
corrections are, in order from largest to smallest (at any scale),
$\ratio{2}$, $\ratio{1}$, $\ratio{3}$.  That such an
ordering should be expected can be seen by examining the RG equation 
for the difference of two ratios (again, Yukawa couplings 
set to zero),
\begin{eqnarray}
\frac{d}{dt} \left( \frac{M_a}{g_a^2} - \frac{M_b}{g_b^2} \right) &=&
    \frac{2}{(16 \pi^2)^2} \sum_c \left( B_{ac}^{(2)} - B_{bc}^{(2)} \right)
    g_c^2 M_c
\end{eqnarray}
where the differences are
\begin{eqnarray*}
B_{2c}^{(2)} - B_{1c}^{(2)} &=& ( -\textfrac{154}{25}, \textfrac{98}{5},
                                  \textfrac{32}{5} ) \\
B_{1c}^{(2)} - B_{3c}^{(2)} &=& ( \textfrac{144}{25}, -\textfrac{18}{5}, 
                                  \textfrac{18}{5} )
\end{eqnarray*}
using the two-loop coefficients given in Appendix~\ref{RGE-app}.
Since the difference between the two-loop coefficients is positive
for the dominant $g_3^2 M_3$ term, the RG equations for the 
difference of the ratios $(\ratio{2} - \ratio{1})$ and
$(\ratio{1} - \ratio{3})$ are also positive,
and thus $\ratio{2}$ obtains the largest two-loop correction,
followed by $\ratio{1}$ and $\ratio{3}$.
This ordering holds if $g_2^2 M_2 \lsim g_3^2 M_3$, and   
it would only be if $g_2^2 M_2 \gg g_3^2 M_3$ and/or 
$g_1^2 M_1 \gg g_2^2 M_2, g_3^2 M_3$ that one would expect 
the ordering of the size of the two-loop corrections to be 
different.

The next stage is to restore the two-loop terms proportional to
the Yukawa couplings.  In the RG equation for the ratio $\ratio{a}$,
each Yukawa coupling (squared) is multiplied by the associated 
soft breaking term, the scalar trilinear coupling, which provides 
the mass scale.  Unlike the ``pure gaugino'' terms, additional 
scales independent of the gaugino masses can partially determine
the RG evolution.  Potentially, one new scale is introduced
for every nonzero Yukawa coupling.  However, we will assume
that the $3 \times 3$ Yukawa coupling matrices $\mathbf{Y}_u$,
$\mathbf{Y}_d$, and $\mathbf{Y}_e$ 
(in flavor space) can be reduced to the dominant third generation
couplings $Y_t$, $Y_b$, and $Y_\tau$, and consequently the only 
relevant scalar trilinear couplings are $A_t$, $A_b$, and $A_\tau$.  
The RG equations for the Yukawa couplings and scalar trilinear
couplings are given in Appendix~\ref{Yukawa-trilinear-app}.
The size of the terms proportional to the Yukawa couplings
depend on the competition between terms proportional $g^2 M$ 
versus those proportional to $Y^2 A$.  It is therefore important 
to recognize that the two-loop corrections to $\ratio{a}$ need not 
be proportional to (or suppressed by) a gauge coupling squared, 
but instead a Yukawa coupling squared.

The Yukawa couplings are extracted at the weak scale from
the fermion mass $m(m)$, but they also depend on $\tan\beta$, 
the ratio of vacuum expectation values of the two neutral Higgs 
doublet scalar fields.  The perturbative lower and upper limits on 
$\tan\beta$ can be found by successfully extrapolating the 
Yukawa couplings up to the high scale $Q_{\mathrm{high}}$
without encountering a Landau pole.  We will be considering 
three cases in the following: one case each of 
low and high $\tan\beta$, and in subsequent discussion
we usually take an intermediate $\tan\beta$ value.

In the low and intermediate $\tan\beta$ cases, we can
safely neglect the terms proportional to $Y_b$ and $Y_\tau$.
If we once again use the initial condition Eq.~(\ref{init-cond-eq}),
normalizing to $M_3/\alpha_3$ at $Q = 1$~TeV, then there
are only two additional parameters, $A_t$ and $\tan\beta$.
We could just as easily substitute $\tan\beta$ for
$Y_t(Q_{\mathrm{high}})$, which is probably a better calculational 
input parameter since it is not sensitively dependent on initial conditions,
thresholds, and the loop order of the Yukawa RG equations, etc.
Of course to precisely translate a $Y_t(Q_{\mathrm{high}})$ back 
into a $\tan\beta$, one must treat the above carefully.
We will provide approximate $\tan\beta$ values corresponding
to particular choices of $Y_t(Q_{\mathrm{high}})$.

\begin{figure}[!t]
\centering
\epsfxsize=4.0in
\hspace*{0in}
\epsffile{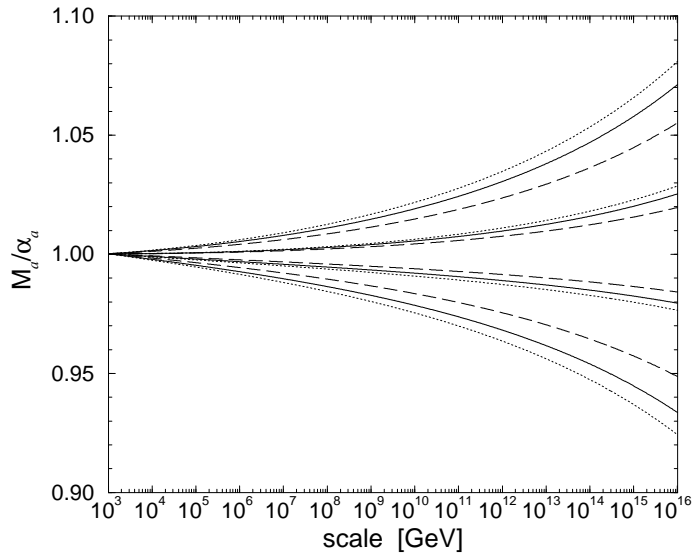}
\caption{Evolution of $\ratio{a}$ as a function of scale, for
$Y_t(10^{16} \; \mathrm{GeV}) = 1$, and 
$A_t = M_3, 0, -M_3, -2 M_3$ at $Q = 1$~TeV for each triplet 
of (solid, dotted, dashed) lines from top to bottom.  
As in Fig.~\ref{case1-fig}, the solid, dotted, and dashed lines 
correspond to $a=1, 2, 3$
respectively.  The normalization differs from Fig.~\ref{case1-fig}
in that the ratio $\ratio{a}$ with Yukawa couplings included is 
normalized against the ratio without Yukawa couplings (identical to
the curves in Fig.~\ref{case1-fig}).}
\label{case2-fig}
\end{figure}

Figure~\ref{case2-fig} shows the effect of a nonzero Yukawa
coupling (and scalar trilinear coupling), with
$Y_t(10^{16} \; \mathrm{GeV}) = 1$, corresponding to
$\tan\beta \approx 2.1$ using one-loop evolution of the
top Yukawa coupling.  [Two-loop corrections to the top Yukawa
coupling typically alter the value of $\tan\beta$ by about $-0.05$
to maintain $Y_t(10^{16} \; \mathrm{GeV}) = 1$.]
The results for $\ratio{a}$ were normalized by
dividing by the ratio $\ratio{a}$ extracted from an equivalent
model with the Yukawa couplings set to zero (i.e., the results
shown in Fig.~\ref{case1-fig}).
The effects of the Yukawa coupling terms are therefore 
evident separately from the pure gaugino terms that enter at two-loop.  
Since it is possible for the gaugino masses and gauge couplings 
to be established experimentally without any detailed knowledge 
of the scalar trilinear couplings, the two-loop 
corrections in Fig.~\ref{case1-fig} are calculable and
do not pose a fundamental theoretical uncertainty in extrapolating
to the high scale (unlike the scalar trilinear couplings).
This provides additional motivation to normalize against the
curves in Fig.~\ref{case1-fig}.

It is clear from Fig.~\ref{case2-fig} that different choices 
for $A_t(1 \; \mathrm{TeV})$ affect the two-loop
running at the level of nearly $10\%$, if a Yukawa coupling
is $\mathcal{O}(1)$.  In the cases 
where $A_t > 0$ and $A_t < - 2 M_3$, 
the scalar trilinear coupling runs to values at the high scale
that are more than a factor of 10 larger than $M_3(1 \; \mathrm{TeV})$.
These input parameters may pose serious problems related to 
fine-tuning (squark masses will be driven to similarly large values), 
and may also be untenable if a charge or color breaking vacuum
is encountered at the weak scale.
To see how large the Yukawa coupling and trilinear scalar coupling
need to be to get a significant correction, 
consider simply the competition between the top Yukawa term 
$C_a^u Y_t^2 A_t$, and the gluino mass term $B_{a3}^{(2)} g_3^2 M_3$.
One immediately observes that the Yukawa terms suffer from suppression 
in the overall constants: $B_{a3}^{(2)}/C_a^u = \textfrac{44}{13}, 
4, \textfrac{7}{2}$ for $a=1, 2, 3$ in the MSSM\@.  If $g_3$ and $Y_t$ 
are of the same order, then $A_t \sim 4 M_3$ to instigate 
corrections of the same order as those of the pure gaugino terms.

\begin{figure}[!t]
\centering
\epsfxsize=4.0in
\hspace*{0in}
\epsffile{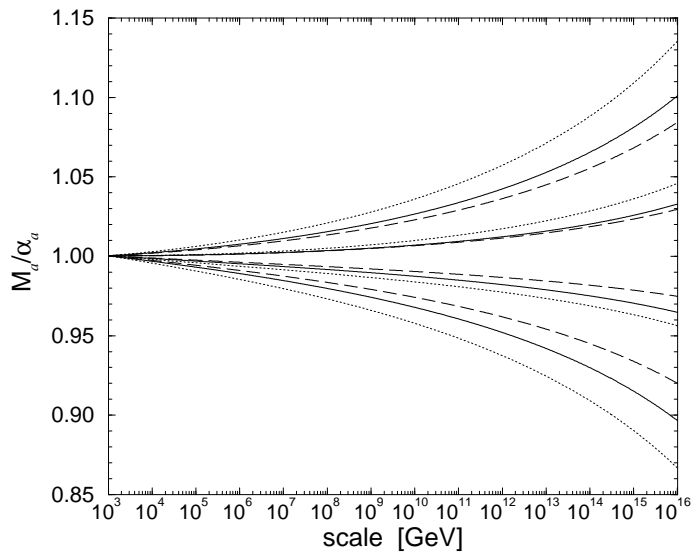}
\caption{Analogous to Fig.~\ref{case2-fig}, with 
$Y_\tau(10^{16} \; \mathrm{GeV}) \approx Y_b(10^{16} \; \mathrm{GeV}) = 1$, 
and the choices $A_t = A_b = A_\tau = M_3, 0, -M_3, -2 M_3$ for each
triplet of lines from top to bottom.}
\label{case3-fig}
\end{figure}

In the case with large $\tan\beta$, both $Y_b$ and $Y_\tau$
evolve to large values, and thus all Yukawa coupling
terms must be included.  In Fig.~\ref{case3-fig} we show
the ratios $\ratio{a}$ normalized as in Fig.~\ref{case2-fig},
with $Y_\tau(10^{16} \; \mathrm{GeV}) \approx 
Y_b(10^{16} \; \mathrm{GeV}) = 1$
corresponding to $\tan\beta \approx 55$, and the choices
$A_t = A_b = A_\tau = M_3, 0, -M_3, -2 M_3$ at $1$~TeV\@.  The size
of the correction is somewhat larger than in Fig.~\ref{case2-fig},
up to nearly $15\%$, mainly because all three Yukawa terms are 
constructively contributing to the $\beta$-function.

Up to now we have been considering the evolution assuming
the initial condition Eq.~(\ref{init-cond-eq}).  Although
the simplest supergravity models and gauge-mediated models 
suggest such a relation should hold near the weak scale, 
it is nevertheless prudent to study other alternatives
(four quite different examples of models with that are not 
expected to have ``unified'' gaugino masses can be 
found in Refs.~\cite{AKKMMboth, AHCM, KoldaMR, lightgluino}).  
Consider a model with gaugino masses $M_a' \equiv k_a M_a$,
such that
\begin{equation}
\frac{k_1 M_1}{\alpha_1} \; \not= \; \frac{k_2 M_2}{\alpha_2}
   \; \not= \; \frac{k_3 M_3}{\alpha_3} \; 
\label{k-init-cond-eq}
\end{equation}
while Eq.~(\ref{init-cond-eq}) holds for $M_a$.  In essence,
we are considering two distinct models with different boundary
conditions on the gaugino masses at the weak scale.
The scaling factors $k_a$ relate a model that does respect
the initial condition, Eq.~(\ref{init-cond-eq}), to one
that does not.  If $M_a' = k_a M_a$ is to hold for 
all scales, then the scaling factors must depend on the 
renormalization scale.  The renormalization group equation 
for the scaling factors can be obtained from the RG equations
for the gaugino masses and gauge couplings,
\begin{eqnarray}
\frac{d}{dt} k_a &=& \frac{2 g_a^2}{(16 \pi^2)^2} 
    \left[ \sum_{b} B_{ab}^{(2)} g_b^2 \frac{M_b}{M_a} (k_b - k_a) 
           + \sum_x C_a^x \frac{A_x}{M_a} \left( \frac{A'_x}{A_x} - k_a 
           \right) \tr (Y_x^\dagger Y_x) \right]
\label{k-full-RGE-eq}
\end{eqnarray}
which represents the evolution of the fraction
\begin{equation}
\frac{M_a'}{M_a} \; = \; \frac{M_a'/\alpha_a}{M_a/\alpha_a} \; .
\end{equation}
If we take only one $k_a$ to be not equal to one, 
and approximate $A'_x/A_x \sim 1$,
Eq.~(\ref{k-full-RGE-eq}) becomes
\begin{eqnarray}
\frac{d}{dt} k_a &=& (1 - k_a) \frac{2 g_a^2}{(16 \pi^2)^2} 
    \left[ \sum_{b; \; b \not= a} B_{ab}^{(2)} g_b^2 \frac{M_b}{M_a} 
           + \sum_x C_a^x \frac{A_x}{M_a} \tr (Y_x^\dagger Y_x) \right] \; .
\label{k-RGE-eq}
\end{eqnarray}
Technically, this is only an approximate RG equation that
works well for $k_1$ or $k_2$, and to a lesser extent $k_3$.
This is because RG equations for the scalar trilinear couplings 
depend on the gaugino masses, and we have assumed
the scalar trilinear couplings are identical for both models 
at all scales to construct Eq.~(\ref{k-RGE-eq}).  However, 
typically the correction is not very large, and thus 
Eq.~(\ref{k-RGE-eq}) is a reasonable approximation to
the ``true'' ratio $(M_a'/\alpha_a)/(M_a/\alpha_a)$.

By comparing the two models using the scaling factors $k_a$,
some analytical insight can be gained into the effect of
the nonstandard initial condition in Eq.~(\ref{k-init-cond-eq})
on the resulting high scale values.
The RG equation, Eq.~(\ref{k-RGE-eq}), indicates that
$k_a$ will tend toward $1$ with increasing scale if $A > 0$
or if simply the pure gaugino terms (positively) dominate over 
the Yukawa terms.  To gain further insight into Eq.~(\ref{k-RGE-eq}),
we can approximately solve for $k_a$ by substituting ``average''
values for the quantities dependent on the renormalization scale,
giving the constant $Z_k$ which we schematically write as
\begin{eqnarray}
Z_k &=& \overline{g}_a^2
    \left[ \sum_{b; \; b \not= a} B_{ab}^{(2)} \overline{g}_b^2 
           \frac{\overline{M}_b}{\overline{M}_a} + \sum_x C_a^x 
           \frac{\overline{A}_x}{\overline{M}_a} 
           \tr (\overline{Y}_x^\dagger \overline{Y}_x) \right] \; .
\end{eqnarray}
Then the solution to Eq.~(\ref{k-RGE-eq}) becomes
\begin{eqnarray}
k_a(Q_f) &=& 1 + \left( k_a(Q_i) - 1 \right) 
   \left( \frac{Q_i}{Q_f} \right)^{\frac{Z_k}{128 \pi^4}} \; .
\end{eqnarray}
With sufficiently small $Q_i/Q_f$ and $Z_k > 0$, it is obvious 
that $k_a$ converges to one as the scale is increased.  
Alternatively, with sufficiently small $Q_i/Q_f$ and $Z_k < 0$, 
$k_a$ will diverge either toward very small or negative values 
(if initially $k_a < 1$) or very large values (if initially $k_a > 1$).
The mundane, but more likely possibility is that
\begin{eqnarray}
\left( \frac{Q_i}{Q_f} \right)^{\frac{Z_k}{128 \pi^4}} &\sim& 1
\end{eqnarray}
and so the initial conditions in Eq.~(\ref{k-init-cond-eq}) are 
preserved under RG evolution.  For $Q_i/Q_f = 10^{-13}$, we find
that $Z_k$ must be
\begin{eqnarray}
Z_k \; \gsim \; ( 20, 100 ) & \quad \mathrm{for} \quad & 
    \left( \frac{Q_i}{Q_f} \right)^{\frac{Z_k}{128 \pi^4}} \; \lsim \;
    (0.95, 0.8) \; ,
\end{eqnarray}
and thus significant deviations to the initial conditions are not
expected after running from the weak scale to the high scale.
We have also confirmed this using numerical calculations.  
However, if $Z_k$ were very large and positive
(a numerical value of several hundred), then $k_a$ will
rapidly evolve toward $1$.  To obtain a very large $Z_k$,
one is faced with large two-loop corrections to the RG
equations.  To be sure perturbation theory is still valid,
one should go to higher orders and check that the three-loop terms
are indeed smaller than the two-loop terms.
In Ref.~\cite{KoldaMR} this was done for models with 
semi-perturbative unification, finding the interesting result that
the initial condition on the gaugino mass ratios is not maintained
under RG evolution, consistent with the above discussion.

\begin{figure}[!t]
\centering
\epsfxsize=4.0in
\hspace*{0in}
\epsffile{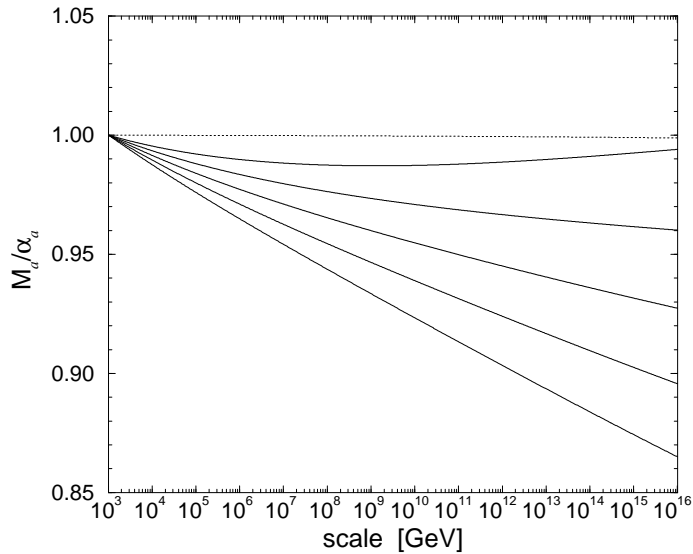}
\caption{The ratio $\ratio{a}$ is plotted versus scale, 
for a model with $\tan\beta = 5$ and $(-\ratio{1}) = \ratio{2} = 
\ratio{3}$ at $1$~TeV.  The curves are normalized by dividing
by a model with the identical initial conditions, except
the sign of $M_1$ is positive.  The set of solid lines correspond 
to the normalized $(-\ratio{1})$, with $A_x(1 \; \mathrm{TeV}) = 
2|M_3|, |M_3|, 0, -|M_3|, -2|M_3|$ from bottom to top.
The dotted line corresponds to the normalized ratios 
$\ratio{2}$ and $\ratio{3}$, which are independent of $A_x$.}
\label{M1-neg-fig}
\end{figure}

Another possibility that can be examined with the formalism
above is to allow $M_1$ and/or $M_3$ to be negative.  This is
possible because in general there can be nonzero phases associated 
with $M_1$ and $M_3$ (the phase of $M_2$ can be chosen to vanish).
We limit ourselves to gaugino masses that are real, 
so that we can concentrate on merely a few different sign 
possibilities.  To examine the impact of a particular sign
choice on the RG evolution, assume the initial condition
\begin{equation}
\pm \frac{M_1}{\alpha_1} \; = \; \frac{M_2}{\alpha_2} \; = \; 
    \pm \frac{M_3}{\alpha_3} \; ,
\label{sign-init-cond-eq}
\end{equation}
where we will examine the two cases $M_1 < 0$, $M_2, M_3 > 0$
and $M_3 < 0$, $M_1, M_2 > 0$.  In Fig.~\ref{M1-neg-fig}
we plot the ratio of $\ratio{a}$ in a model with $M_1 < 0$
normalized to an equivalent model with $M_1 > 0$, and all
other initial conditions the same (we take the intermediate
value $\tan\beta = 5$ for the purposes of this example).  
Since the dominant
term in the evolution is $g_3^2 M_3$, it is not surprising
that $\ratio{2}$ and $\ratio{3}$ are virtually unaffected 
by the sign choice of $M_1$.  However, $\ratio{1}$ is 
significantly affected.  Setting $k_1(1 \; \mathrm{TeV}) = -1$,
one finds that Eq.~(\ref{k-RGE-eq}) captures essentially all 
of the difference between a model with and without a negative 
$M_1$.  The size of the effect after running $13$ orders
of magnitude in scale varies depending on the competition
between the pure gauge terms and the Yukawa terms in the RG equation. 
In the case where $A_x < 0$, for example, there is a partial 
cancellation between the pure gaugino terms (dominated by
$g_3^2 M_3 > 0$) and the Yukawa terms, thus the RG evolution 
is not as pronounced.

\begin{figure}[!t]
\centering
\epsfxsize=4.0in
\hspace*{0in}
\epsffile{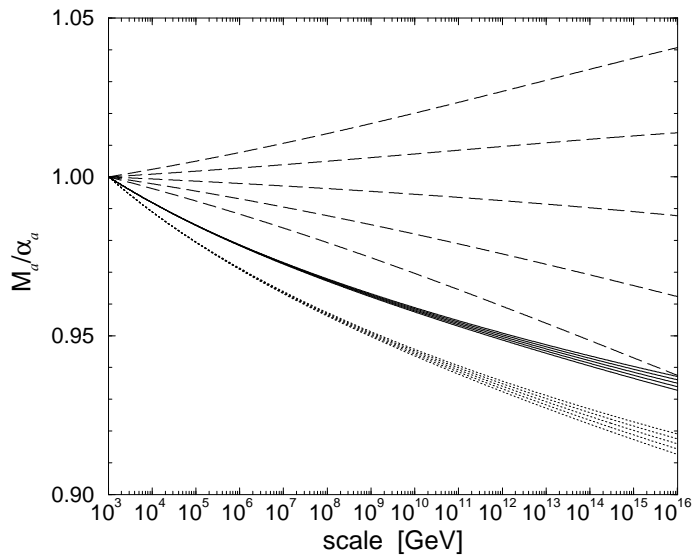}
\caption{Analogous to Fig.~\ref{M1-neg-fig}, with 
$\ratio{1} = \ratio{2} = (-\ratio{3})$ at $1$~TeV.
The set of solid, dotted, and dashed lines correspond to the 
normalized $\ratio{1}, \ratio{2}, (-\ratio{3})$, with 
$A_x(1 \; \mathrm{TeV}) = 2|M_3|, |M_3|, 0, -|M_3|, -2|M_3|$
from top to bottom for $(-\ratio{3})$, and from bottom to top
for the other ratios.}
\label{M3-neg-fig}
\end{figure}

In the case where $M_3 < 0$ and $M_1, M_2 > 0$, and the initial condition
Eq.~(\ref{sign-init-cond-eq}) holds, there is a more dramatic
effect on the RG evolution of all the gaugino masses
due to the dominance of the $g_3^2 M_3$ term.  In 
Fig.~\ref{M3-neg-fig} we illustrate the RG evolution
for the same initial conditions as in Fig.~\ref{M1-neg-fig}, 
but with $M_3 < 0$ instead of $M_1 < 0$.  We observe that
the evolution of $(-\ratio{3})$ is \emph{not} as significant 
as the evolution of $(-\ratio{1})$ (for this example), 
mainly because the $g_3^2 M_3$ term is absent in Eq.~(\ref{k-RGE-eq})
for $a=3$.
(An equivalent way of thinking about this is that the $g_3^2 M_3$
dominates the RG evolution for $\ratio{3}$, which implies an 
approximate $M_3$ sign independence, see Eq.~(\ref{ratio-RGE-eq}).)
This also implies the Yukawa terms dominate over the pure 
gaugino terms in Eq.~(\ref{k-RGE-eq}), giving the logarithmically
increasing but well-defined separation between the 
different $(-M_3/\alpha_3)$ curves as $A_x$ 
is increased.  However, we should point out that Eq.~(\ref{k-RGE-eq})
only approximately accounts for the RG evolution of the
ratio; this is because, as noted above, Eq.~(\ref{k-RGE-eq})
does not account for the $M_a$ sign dependent part of the
evolution of $A_x$.

\subsection{Weak scale thresholds}
\label{weak-thresholds-sec}
\indent

We have up to now avoided the issue of weak scale threshold
corrections by judiciously choosing the initial conditions, 
including the gaugino masses, scalar trilinear couplings, etc., 
to hold at a scale \emph{above} all weak scale thresholds.
The motivation for this is that one can always
match a weak scale theory including full threshold corrections
to the initial conditions we gave in the examples above.  
In practice, this matching can become rather 
complicated~\cite{PP, BMP, PBMZ}
if one demands a high level of precision.  
However, it is important to understand the origin of
the uncertainties associated with weak scale thresholds, as well as
recognizing that, for example, measured (pole) masses must be translated 
into renormalized masses, and the corrections can be large (especially
for the gluino~\cite{MVGaugino, PP, PBMZ}).  Note that we have implicitly 
assumed the running gaugino masses are specified in the scheme 
appropriate for supersymmetry, namely dimensional reduction with modified 
minimal subtraction ($\overline{\mathrm{DR}}$)~\cite{DRED}.

Threshold corrections arise from decoupling heavy particles
from the spectrum by creating an effective theory without
the heavy degrees of freedom that matches near the scale 
of the heavy particles.  In weak scale supersymmetric 
theories one is interested in decoupling sparticles that
are heavier than $M_Z$, the customary choice of scale for 
starting RG evolution since the gauge couplings are 
very well measured by LEP experiments.  There are two methods 
for handling threshold corrections:
(1) ``decoupling method'', where heavy particles are decoupled 
by altering the $\beta$-functions at the mass scale of the particles, 
and (2) ``nondecoupling method'', where fully
supersymmetric $\beta$-functions are retained and the effects
of heavy particles are ``resummed'' as corrections
in the conversion from the measured to running quantity.
Strictly speaking, the second method is best since 
full logarithmic and finite corrections with arbitrary mixing 
can be incorporated (although there are sometimes 
ways of re-incorporating finite corrections using the 
decoupling method).
However, explicitly resumming logarithms using a nondecoupling
method is really only useful when the scale of 
the decoupled particles is near the scale of the conversion 
from measured to running quantity, otherwise the first method
should be used.  For completeness, we give formulae to compute
the one-loop $\beta$-function decoupling for the weak scale
threshold corrections in Appendix~\ref{thresholds-app}.

It is useful to first recall that the RG equation for
the ratio $\ratio{a}$ is independent of one-loop threshold
corrections implemented as changes in the one-loop $\beta$-function
coefficients.  This is obvious from the RG equation,
Eq.~(\ref{ratio-RGE-eq}), since it is independent of $B_a^{(1)}$.
Threshold corrections at two-loop would, however, change the 
coefficients for the ratio's RG equation.  But, weak scale threshold 
corrections at two-loop are negligible when we are working 
with two-loop RG evolution equations~\cite{Hall}, since no large
logarithm develops (precisely because we demand the scale of 
the superpartner masses be near the weak scale).
There are residual indirect effects from one-loop thresholds
that arise because of the two-loop terms proportional 
to the gauge couplings, but in virtually all instances such
corrections can be neglected.

Take the simplest case, the gluino mass~\cite{MVGaugino, PP, PBMZ}.  
It is easy to show that in the limit 
$m_q \ll m_{\tilde{g}} \ll m_{\tilde{q}}$ the logarithmic 
corrections scale with $\log m_{\tilde{q}}^2/m_{\tilde{g}}^2$
when the gluino (running) mass is evaluated at a 
scale $Q = m_{\tilde{g}}$.  The coefficients of
the log terms precisely match the shifts in the one-loop $\beta$-functions
(see Appendix~\ref{thresholds-app}).  However, 
nonzero quark masses bring additional corrections, which can only
be partially taken into account using a decoupling method.
Furthermore, the corrections to the weak gaugino masses are
complicated due to the mixings inherent in the resultant charginos
and neutralinos~\cite{PP, PBMZ}.  Luckily, incorporating
the logarithmic corrections to the gluino mass, and incorporating
approximate logarithmic corrections (neglecting mixings)
in the weak gaugino masses is usually sufficient for
most purposes.  Since we specified running masses throughout
the previous discussions, it was not necessary to explicitly
calculate the corrections from translating the pole mass into 
the running $\overline{\mathrm{DR}}$ mass.  However, experiments
will ultimately measure the pole mass, and so these finite corrections 
must be taken into account.  In general, extracting parameters 
from experiment will require a careful analysis by incorporating 
both finite corrections and logarithmic corrections (probably 
using a nondecoupling method as in Ref.~\cite{PBMZ}).

\subsection{Superoblique corrections}
\indent

There is further class of weak scale 
corrections~\cite{FMPT, NFT, HN, CFP, Randall, NPY, KRS} that 
result when supermultiplets are widely split in 
mass.  These so-called
``superoblique'' corrections have much in common
with oblique corrections of the standard 
model~\cite{CFP, Randall}, particularly because
they do not decouple for scales smaller than the
heaviest sparticle mass.  The manifestation of 
superoblique corrections is the violation the 
relation $g = \tilde{g}$, where $g$ is the coupling
of gauge bosons to fermions and scalars, and $\tilde{g}$
is the coupling of gauginos to a fermion and
its scalar partner.  Ordinarily supersymmetry enforces
this relation to all orders (in a dimensional 
reduction scheme~\cite{MVGaugino}), but since
supersymmetry must be broken near the weak scale,
differing corrections to $g$ and $\tilde{g}$ are expected.

Differentiating between the gauge and gaugino coupling
suddenly begs the question of which one ought to be
used in the ratio $\ratio{a}$.  To address this,
there are four cases to be distinguished.  To simplify
the discussion, consider the effects of a single supermultiplet 
consisting of a scalar $\tilde{q}$ and its fermion partner $q$
with masses $m_{\tilde{q}}$ and $m_q$ respectively.
(The extension to multiple multiplets and specialization
to a particular gaugino is trivial.)
The four cases are (1) $m_{\tilde{q}}, m_q \ll M$,
(2) $M \ll m_{\tilde{q}}, m_q$, 
(3) $m_q \ll M \ll m_{\tilde{q}}$, and 
(4) $m_{\tilde{q}} \ll M \ll m_q$, 
where $M$ is the gaugino mass.  The first case is trivial; 
it is easy to show that there is no distinction between
$g$ and $\tilde{g}$ when all matter multiplets have
masses well below that of the gaugino mass.  The second
case is nontrivial, and has been studied particularly
for the messenger sector of gauge-mediated models~\cite{CFP, KRS}.  
Even in cases when the lightest messenger
scalar is $1/10$ the mass of the messenger fermion, 
one finds the fractional difference $(g - \tilde{g})/g \lsim 10^{-4}$.
It is really only the case where either a fermion is very
heavy and scalars are light, or scalars are heavy and the fermion
is light, when superoblique corrections are significant.  
Motivated by the current experimental bounds on the masses
of the scalar partners that require $m_q \ll m_{\tilde{q}}$ for
the MSSM matter multiplets\footnote{With the possible exception
of the scalar partner to the top quark.},
we will focus on the latter possibility.

\begin{figure}
\begin{picture}(457,55)(0,0)
  \Line( 40, 20 )( 180, 20 )
  \Photon( 40, 20 )( 80, 20 ){3}{4}
  \Photon( 140, 20 )( 180, 20 ){3}{4}
  \DashCArc( 110, 20 )( 30, 0, 180 ){3}
  \Text( 110, 4 )[c]{(a)}
  \Line( 270, 20 )( 410, 20 )
  \Photon( 270, 20 )( 410, 20 ){3}{14}
  \PhotonArc( 340, 20 )( 30, 0, 180 ){3}{8}
  \Text( 340, 4 )[c]{(b)}
\end{picture}
\caption{One-loop self-energy corrections to the gaugino
mass from (a) sfermion-fermion contributions, and (b) gauge boson
contributions.}
\label{gaugino-fig}
\end{figure}
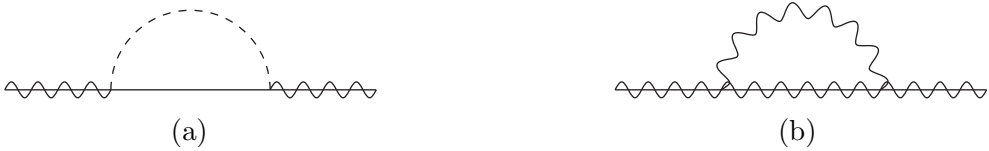

The one-loop corrections to the self-energy of
the gaugino are shown in Fig.~\ref{gaugino-fig}.
We are interested in the sfermion-fermion one-loop corrections
that contain the superoblique corrections.  The self-energy
due to the sfermion-fermion diagram in Fig.~\ref{gaugino-fig}(a) is
\begin{eqnarray}
\Sigma(\slashchar{p}) &=& \frac{2 g^2 S(q)}{16 \pi^2} \slashchar{p}
    \int_0^1 dx \> (1-x) \ln \frac{x m_q^2 + (1-x) m_{\tilde{q}}^2 
    - x (1-x) p^2}{\mu^2} \; ,
\end{eqnarray}
where $p$ in the momentum of the gaugino, $\mu$ is the renormalization 
scale, and we have performed $\overline{\mathrm{DR}}$ subtraction 
of the infinite piece.  The Dynkin index $S(q)$ is defined by
$S(q) \delta^{ab} \equiv \mathrm{tr}_q (t^a t^b)$ in a normalization
where $S(q) = \textfrac{1}{2}$ for the fundamental of $SU(N)$,
and $S(q) = \textfrac{3}{5} (Y/2)^2$ for $U(1)_Y$ in the GUT
normalization.  The correction to the running gaugino mass 
due to the superoblique corrections can be written as
\begin{eqnarray}
M(M) &=& M(M)_0
    \left( 1 + \frac{\alpha}{4\pi} \sum_q A_q \right) \;. 
\label{mass-correction-eq}
\end{eqnarray}
where $M(M)$ and $M(M)_0$ are the running $\overline{\mathrm{DR}}$
gaugino masses with (without) one-loop superoblique corrections 
applied.  The function $A_q$ is
\begin{eqnarray}
A_q &=& 2 S(q) \int_0^1 dx \> (1-x) \ln \frac{x m_q^2 + (1-x) m_{\tilde{q}}^2 
    - x (1-x) [M(M)_0]^2}{\mu^2} \; ,
\end{eqnarray}
approximating $\slashchar{p} = M^{\mathrm{pole}} \approx M(M)_0$.
The above expression agrees with the squark-quark one-loop
corrections calculated in Refs.~\cite{MVGaugino, KRS}.
Note that there are additional one-loop finite corrections due 
to gauge boson loops [shown in Fig.~\ref{gaugino-fig}(b)] 
that we do not present, since they can be absorbed into the
translation between $M^{\mathrm{pole}}$ 
and $M(M)$~\cite{YamadaGaugino, MVGaugino}.
The correction to the gaugino coupling arises from
wave-function renormalization $Z_2$ of the gaugino propagator
$i Z_2/(\slashchar{p} - M + i\epsilon)$, where
$Z_2^{-1} = 1 - \textfrac{d}{d \slashchar{p}} \Sigma(\slashchar{p})$
after expanding the denominator of the one-loop propagator 
$i/(\slashchar{p} - M - \Sigma(\slashchar{p}) + i\epsilon)$,
in $(\slashchar{p} - M)$.  The result is a correction to
the gaugino coupling
\begin{eqnarray}
\tilde{g}^2(M) &=& \tilde{g}_0^2(M)
    \left( 1 + \frac{\alpha}{4\pi} \sum_q A_q \right) \; .
\label{coupling-correction-eq}
\end{eqnarray}
using notation analogous to that in Eq.~(\ref{mass-correction-eq}).
(There are additional nonlogarithmic corrections~\cite{KRS} 
to Eq.~(\ref{coupling-correction-eq}), but they decouple 
when $M(M) \ll m_{\tilde{q}}$.)
Thus, the ratio $M_a/\tilde{g}_a^2$ is independent
of superoblique threshold corrections.  This is 
an important result, because it means that using the
coupling $g$ (that is more easily extracted from gauge 
boson interactions) to evolve the ratio $\ratio{a}$ will 
reintroduce one-loop corrections for scales between
$M < Q < m_{\tilde{q}}$.  Alternatively, the uncertainty
in translating $g$ into $\tilde{g}$ due to superoblique
corrections is a fundamental uncertainty in evaluating
the true one-loop invariant $M/\tilde{g}^2$ near the weak scale.  

\begin{figure}[!t]
\centering
\epsfxsize=4.0in
\hspace*{0in}
\epsffile{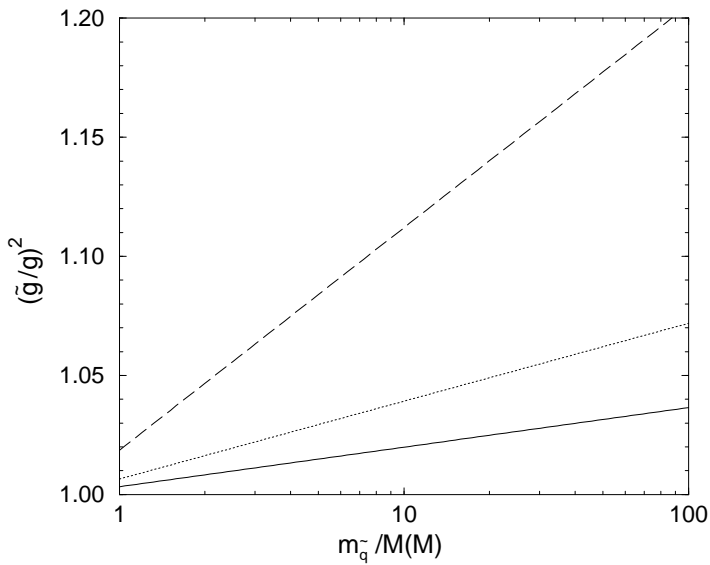}
\caption{An example of superoblique corrections to $M_a/g_a^2$,
due to setting the masses of the first two generations 
of squarks and sleptons to be $m_{\tilde{q}}$.  The ratio
$(\tilde{g}_a/g_a)^2 = (M_a/g_a^2)/(M_a/\tilde{g}_a^2)$, where
the denominator $M_a/\tilde{g}_a^2$ is independent of superoblique
corrections, but the numerator $M_a/g_a^2$ is expected to be
extracted from experiment.  The solid, dotted, and dashed lines
correspond to $a = 1, 2, 3$.  We assumed the initial 
condition, Eq.~(\ref{init-cond-eq}), with $M_1(M_1) = 100$~GeV,
and evaluated the correction at the scale $M_a(M_a)$.}
\label{superoblique-fig}
\end{figure}
The size of the correction (including finite corrections) was
given in Ref.~\cite{KRS}, and takes the form
\begin{eqnarray}
\left( \frac{\tilde{g}}{g} \right)^2 &=&
    1 + \sum_q S(q) \frac{\alpha}{6 \pi} 
    \left( \ln \frac{m_{\tilde{q}}^2}{Q^2} + \frac{11}{12} \right)
\end{eqnarray}
where the sum is over all scalars $\tilde{q}$ transforming under the
gauge group associated with $g$ (or $\tilde{g}$) with Dynkin index $S(q)$.
Again, we are considering the case where the scalar partners are
much heavier than the fermions.  Of course we could have also 
considered the opposite case (see Refs.~\cite{CFP, KRS}),
but there is motivation, for example, to set some or all of 
the scalar partners of the first two generations in the MSSM 
to be very heavy, of order $\sim 20$~TeV or 
so~\cite{DG, anomalousU1, nelson-heavy-squark}.  These models
avoid flavor changing neutral current (FCNC) constraints from
experiment (that dominantly restrict the first two generations)
by pushing the mass scale of the exchanged scalars sufficiently 
high so that universality and alignment are unnecessary.
In these ``2--1'' models the correction becomes~\cite{CFP, KRS}
\begin{eqnarray}
\left( \frac{\tilde{g}_a}{g_a} \right)^2 &=&
    1 + \frac{2 \alpha_a}{3 \pi} 
    \left( \ln \frac{m_{\tilde{q}}^2}{Q^2} + \frac{11}{12} \right) \; ,
\end{eqnarray}
where we have assumed the scalar partners of the first 
two generations are degenerate with a mass $m_{\tilde{q}}$.  
If we further assume the initial condition, Eq.~(\ref{init-cond-eq}),
then the size of the correction at $M(M)$ can be evaluated,
as shown in Fig.~\ref{superoblique-fig}.
It is clear that using the ratio $M_a/g_a^2$ (uncompensated for
the superoblique corrections) implies corrections of order
a few to tens of percent, depending on the gaugino involved 
and the separation of scales.

\section{Two-loop effects beyond the MSSM}
\label{beyond-sec}
\indent

Throughout the previous section various effects were 
discussed that could change the evolution of
$\ratio{a}$ in the MSSM\@.  Extensions of the MSSM
could also easily give corrections to $\ratio{a}$, 
usually through modifications to the two-loop $\beta$-function
coefficients.  In the following, two classes of well-motivated 
extensions of the MSSM will be explored.  The first postulates 
an extra $U(1)'$ symmetry, using both a ``minimal'' $U(1)'$ model
and an $E_6$ model as examples.  The second extension postulates 
extra matter between the weak scale and the high scale.  

\subsection{Extra $U(1)'$}
\indent

The group structure of the MSSM can be easily extended
to include an extra $U(1)$ group.  There are many motivations
for such an extension~\cite{CLreview}, such as
a solution to the $\mu$-problem, string or GUT breaking
to the MSSM plus an additional $U(1)$, etc.
However, enlarging the group structure of the MSSM does 
not come without its own subtleties.  The most obvious difficulty 
is to assign $U(1)'$ charge such that all the anomalies are 
canceled; this is, in general, a difficult problem.  
In Appendix~\ref{anomalies-app} we present the
conditions on the multiplicities and charges of matter transforming
under a $U(1)'$ that cancel the anomalies.  Although 
the $U(1)_Y$ anomaly conditions were not explicitly stated,
cancellation is guaranteed if, for example, matter is added in
complete representations of $SU(5)$ or is added in
vectorlike pairs transforming under the MSSM gauge group.  
It was shown in Ref.~\cite{Langacker} that
a simple model can be constructed with $U(1)'$ charges to 
one generation (only) of the MSSM matter that is nonanomalous, 
and this will serve as our ``minimal'' $U(1)$ example.
We will also examine an $E_6$ model with three complete 
generations of $\mathbf{27}$s, which is also well known 
to be nonanomalous.
We will not attempt to survey all possible extensions that 
include an extra $U(1)$, but instead consider mainly just these
two classes of models.  However, the particular breaking
pattern of a GUT group, such as $E_6$, can change the results 
depending on the scale of the breaking and whether other
group structures exist at intermediate scales.  We will
briefly comment on this in Sec.~\ref{GUT-breaking}.
Nevertheless, we expect that the results would not be significantly 
different if another $U(1)'$ model ansatz were chosen that was 
valid up to near the high scale.

Before we present results for particular models, 
it is useful to understand the origin of corrections
to the ratio $\ratio{a}$ due to a $U(1)'$ extension.
First, the superpotential is assumed to have the term
\begin{eqnarray}
W &\subset& Y_S S H_1 H_2
\label{superpotential-addition-eq}
\end{eqnarray}
in place of $\mu H_1 H_2$, where the superfield $S$ is a gauge 
singlet under $SU(3)_c \times SU(2)_L \times U(1)_Y$.
This in itself is a modification of the MSSM that could 
be considered separately, but comes with its own set of 
difficulties~\cite{NMSSM}.  We will concentrate on a model
with an extra $U(1)$ that will be broken when $S$ acquires
a vacuum expectation value (vev).  To avoid reintroducing 
the $\mu$-problem into these models, the term $\mu H_1 H_2$ is 
assumed to be forbidden (by a judicious assignment of $U(1)'$ charge).
When $S$ acquires a vev, an effective $\mu = Y_S \langle S \rangle$
is generated, and thus the CP-odd Higgs boson and all neutralinos 
in the MSSM will acquire mass.
The particle content of this minimal extension 
includes a $Z'$, a new gaugino $\lambda'_S$, a new Higgsino 
$\tilde{S}$, a scalar $S$, and a pseudoscalar.  The pseudoscalar 
associated with $S$ is eaten by the $Z'$, giving it mass 
via the Higgs mechanism.  Three new soft terms are introduced
\begin{eqnarray}
-\mathcal{L}_{\mathrm{soft}} &\subset& m_{S}^2 |S|^2 
    + \textfrac{1}{2} (M' \lambda_S' \lambda'_S + h.c.)
    + (Y_S A_S S H_1 H_2 + h.c.)
\end{eqnarray} 
(we take $m_S$, $M'$, and $A_S$ to be real)
to give mass to the new gaugino and the uneaten scalar from 
the MSSM singlet superfield $S$, and a scalar trilinear coupling
associated with the new Yukawa coupling.  The neutralino mass matrix
enlarges to $6 \times 6$, and the neutral CP-even Higgs 
mass matrix enlarges to $3 \times 3$.  A new gauge coupling
$g'$ exists for the $U(1)'$, giving two new couplings
($g'$ and $Y_S$) that enter the two-loop RG equations
for the gauge couplings and gaugino masses.  There is
also a new one-loop invariant $M'/{g'}^2$.
For a general set of $U(1)'$ charges $Q_i$ assigned
to the MSSM multiplets, we have computed\footnote{When
$g_a$ and $M_a$ are specified for a model with an additional $U(1)$, 
the $a=4$ elements correspond to $g'$ and $M'$.}
the two-loop renormalization group equations for all the gauge
couplings ($g_1$, $g_2$, $g_3$, $g'$) and gaugino masses
($M_1$, $M_2$, $M_3$, $M'$) including the effects
of the additional Yukawa coupling $Y_S$.  The results
are presented in Appendix~\ref{RGE-app}.

Unlike the MSSM gauge and Yukawa couplings, the gauge coupling 
$g'$ and the Yukawa coupling $Y_S$ are not determined by
low energy experiments, and are therefore essentially free parameters.
(Of course there will be constraints on these parameters in 
particular models from nonobservation data.)  
The size of the effects of the additional
$U(1)'$ are therefore highly model dependent.  In practice, 
the two-loop corrections are not expected to be arbitrarily 
large unless $g'$ or $Y_S$ approach a Landau pole near the 
high scale.  However, we will only consider models in which 
$g'$ and $Y_S$ can be treated perturbatively throughout the energy
scale of interest, which will limit the size of the effect.

There is an additional parameter when two or more $U(1)$
groups are present in the low energy effective theory.  
The kinetic terms of $U(1)_Y$ can mix with those of $U(1)'$
with a strength that is \emph{a priori} unknown and a free parameter
(again, in particular models there are constraints from 
experiment) \cite{Holdom, delAguila, BKMR, DKMR, BKMRrecent}.
Furthermore, the mixing is not RG invariant if 
\begin{eqnarray}
\tr (Q_a Q_b) \not= 0 \;
\end{eqnarray}
the product of charges of the two groups, $U(1)_a \times U(1)_b$, 
summed over all chiral representations in the theory is nonzero.
Although it would be interesting to know if there are 
consequences of kinetic mixing on the evolution of $\ratio{a}$, 
we do not expect qualitatively different results, 
and therefore we do not consider this further in this paper.
In particular, we will consider extra $U(1)$ models that 
obey $\tr (Y Q) = 0$ and have their kinetic mixing set
(by hand) to zero.

\subsection{Minimal $U(1)'$ model}
\label{minimal-model-sec}
\indent

One of the simplest extensions of the MSSM that includes an
additional $U(1)'$ was extensively studied in Ref.~\cite{Langacker}.
They proposed a superpotential with two Yukawa couplings, 
\begin{eqnarray}
W &=& Y_u Q H_2 u^c + Y_S S H_1 H_2 \; ,
\end{eqnarray}
where the up-type Yukawa $Y_u$ couples only to the third
generation, and the down-type Yukawas $Y_d$ and $Y_e$ are absent
(actually forbidden by $U(1)'$ charge assignments).
The $U(1)'$ charge assignments consistent with the cancellation
of anomalies are given in Appendix~\ref{minimal-app}.
The undetermined parameters of the model in addition to those
in the MSSM are (in our conventions): the product of the $U(1)'$
charge for $H_1$ and the $U(1)'$ gauge coupling $Q_1 g'$, 
the $U(1)'$ gaugino mass $M'$, the Yukawa coupling $Y_S$, 
and the associated scalar trilinear coupling $A_S$.

Since $U(1)$ couplings are not asymptotically free, any additional
$U(1)'$ coupling added to the theory will have the greatest effect on
the RG evolution of other parameters near the high scale.  
In general, it is hard to achieve any
significant correction due solely to a larger $g'$ without
running to a Landau pole very near the high scale.
In the following, we will take the initial condition in 
Eq.~(\ref{init-cond-eq}),
to hold at $1$~TeV, analogous to the analyses done in Sec.~\ref{MSSM-sec}.
Furthermore, $Q_1 g'$ is set to $g_1 (= g_2)$ at the high scale,
$M' = c M_3$ at the weak scale where $c$ is a constant, and 
for now Yukawa contributions to the two-loop RG equations are ignored.
With these assumptions, the correction to $\ratio{a}$ 
is less than $0.004 c$ after evolving $13$ orders of 
magnitude in scale.  Even if the coupling $Q_1 g'$ is increased
to of order one, the correction is smaller than about $0.01 c$.
Unless $M'$ is taken to be considerably larger than $M_3$
(which would be disfavored by naturalness arguments), 
the correction due to ``pure $U(1)'$ gaugino'' contributions
to two-loop running is essentially negligible.

\begin{figure}[!t]
\centerline{
\hfill
\epsfxsize=0.55\textwidth
\epsffile{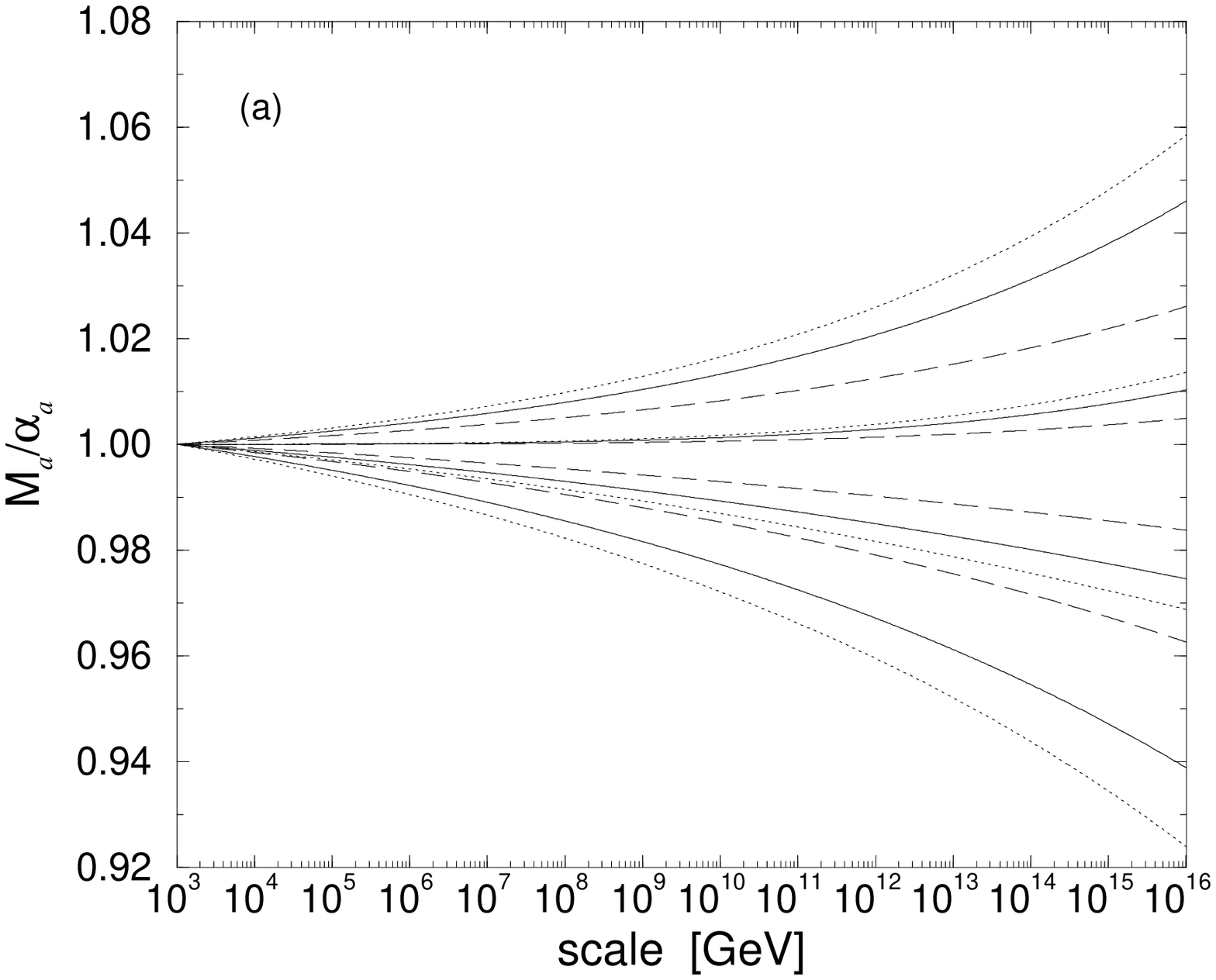}
\hfill
\epsfxsize=0.55\textwidth
\epsffile{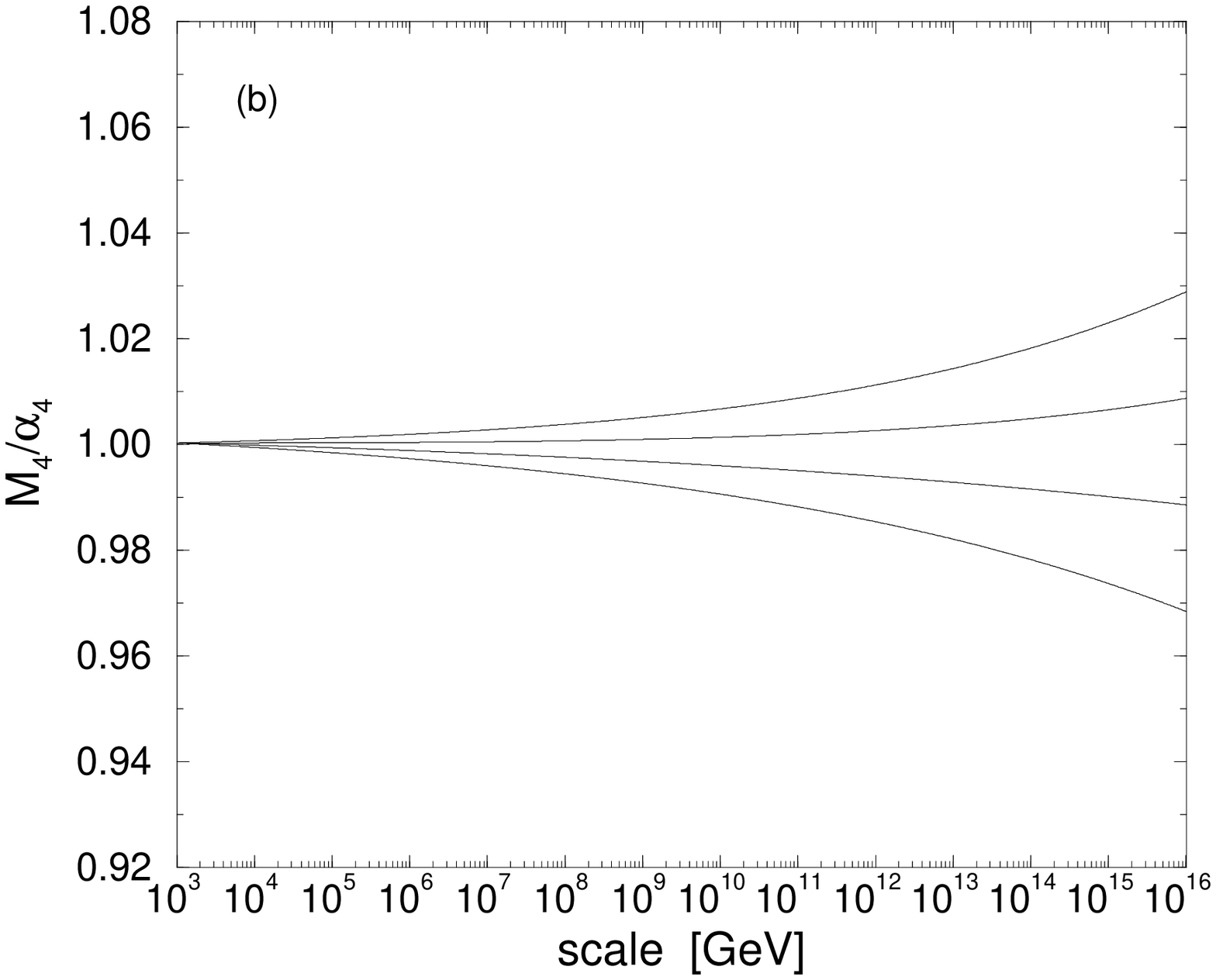}
\hfill }
\caption{Evolution of $\ratio{a}$ as a function of scale, for
$Y_S(10^{16} \; \mathrm{GeV}) = 1$, and 
$A_t = A_S = M_3, 0, -M_3, -2 M_3$ at $Q = 1$~TeV from top to bottom 
for each triplet of lines in (a), and for each line in (b).  
The solid, dotted, and dashed lines correspond to $a=1, 2, 3$
respectively in (a), and the solid lines correspond to $a = 4$ in (b).
The ratio $\ratio{a}$ has been normalized against an equivalent
model with the Yukawa couplings set to zero.}
\label{langacker-fig}
\end{figure}

However, when the Yukawa couplings are nonzero, larger deviations 
can be obtained without soft parameters exceeding of order $M_3$
near the weak scale.  In Fig.~\ref{langacker-fig} we show
the corrections to $\ratio{a}$ arising from a model with 
$Y_S = 1$ at the high scale, and $Y_t$ finite (we set 
$\tan\beta \approx 5$; lower or higher values of $\tan\beta$ 
give corrections analogous to Figs.~\ref{case2-fig} 
and \ref{case3-fig}).  
Again, in complete analogy to
the MSSM studied in Sec.~\ref{MSSM-sec}, large Yukawa
couplings typically imply that the scalar trilinear couplings
run to multi-TeV scale values near the high scale,
which causes the nontrivial correction in $\ratio{a}$.
The size of the corrections are, in order from largest to smallest,
$\ratio{2}$, $\ratio{1}$, $\ratio{4}$, $\ratio{3}$.  The ratio
$\ratio{3}$ receives the smallest correction because the
$\beta$-function coefficient $C_3^S = 0$, and thus it is only 
through $Y_S$ feeding into $Y_u$ that there is any correction 
at all.

\subsection{$\eta$-model of $E_6$}
\label{eta-E6-sec}
\indent

There has been an enormous amount of work studying the 
phenomenology of superstring theories, particularly
the group structures that emerge from compactification.
One distinct possibility is $E_6$~\cite{BDHS, EENZ, 
IM, DAGQ, E6report, Dienes}, 
which can arise from $E_8$ under suitable breaking.  
$E_6$ itself is rank-6, and one commonly studied breaking pattern is 
$E_6 \ra SO(10) \times U(1) \ra SU(5) \times U(1) \times U(1)$.  
Although the effects
of two additional $U(1)$'s could itself be a particularly
interesting possibility to study, we restrict ourselves to 
considering the rank-5 $\eta$-model $SU(5) \times U(1)_\eta$,
and further assume $SU(5)$ breaks near the scale where $E_6$
was broken.  (Changing the scale where the GUT breaks, and 
other possible breaking patterns will be discussed in 
Sec.~\ref{GUT-breaking}.)
The $U(1)_\eta$ is a mixture of the two $U(1)$'s in $E_6$ that
is assumed to survive\footnote{The other
$U(1)$ orthogonal to $U(1)_\eta$ is assumed to break 
at the high scale.  In this section, we neglect effects
of this high scale $U(1)$.} to near the weak scale.
This provides a well-motivated example of a model with an
extra $U(1)$ in which the charge assignments and overall
normalization are determined (by embedding into $E_6$;
see Table~\ref{charge-table} for the charge assignments).

In particular, we consider an $\eta$-model with three
generations of $\mathbf{27}$s, which is the smallest number
of representations that incorporates the MSSM and is 
nonanomalous.  Unlike the minimal 
$U(1)'$ model considered above in Sec.~\ref{minimal-model-sec},
all three generations are charged under the $U(1)'$, and
Yukawa couplings for $b$ and $\tau$ are present.  Although the 
latter is not expected to affect the results of evolving the 
ratios $\ratio{a}$ to the high scale (unless $\tan\beta$ is large), 
it does overcome a limitation of the minimal model.  
Three complete $\mathbf{27}$
representations of $E_6$ include three generations of the following:
MSSM matter ($Q$, $u^c$, $d^c$, $L$, $e^c$), down-type and up-type 
Higgs doublets ($H_1$, $H_2$), two color triplets ($D$, $D^c$), 
a right handed neutrino ($\nu^c$), and a MSSM gauge singlet ($S$).
The superpotential for the model is
\begin{eqnarray}
W &=& Y_t Q H_2 u^c + Y_d Q H_1 d^c + Y_e L H_1 e^c + Y_S S H_1 H_2
    + Y_D S D D^c \; .
\label{E6-superpotential-eq}
\end{eqnarray}
Additional terms are possible, and indeed some are probably 
necessary for a phenomenologically viable model~\cite{ENPZ, E6report}.
However, additional dimension-3 terms in the superpotential 
with small Yukawa couplings will not affect the results 
presented below.
To prevent large tree-level flavor changing neutral
currents, only one set of Higgs ($H_1$, $H_2$, $S$) is assumed 
to couple to the quark and lepton superfields.  To break the
$U(1)_\eta$, give mass to the color triplets, and generate 
an effective $\mu$ term, the scalar component of $S$ is assumed
to acquire a vev.
While the details of the vacuum structure (including 
avoiding possible charge/color breaking minima) are very 
important to be able to construct viable models of $E_6$, 
we will not discuss this further\footnote{The mapping of 
our results onto completely viable weak scale models derived 
from $E_6$ is beyond the scope of this paper.}.
Our intention is to take the matter content of an $E_6$ model with
the Yukawa couplings given in Eq.~(\ref{E6-superpotential-eq}),
and examine the consequences for the evolution of $\ratio{a}$.

The RG equations for a three generation $E_6$ model are
given in Appendix~\ref{RGE-app}, including the two-loop
terms proportional to $Y_S^2$ and $Y_D^2$ in the
gauge coupling and gaugino mass RG equations. 
It is well known that an $E_6$ model with three complete 
$\mathbf{27}$s does not preserve gauge coupling unification,
due to the extra Higgs doublets and color triplets.  
Although it is possible to add extra matter to the 
theory to bring the gauge couplings back into 
alignment at the high scale (see e.g.\ \cite{DVJ, KeithMa, BKMR}), 
we defer a discussion of the effects of extra matter (beyond that
needed to fill a $\mathbf{27}$ of $E_6$) to the next section.

\begin{figure}[!t]
\centerline{
\hfill
\epsfxsize=0.55\textwidth
\epsffile{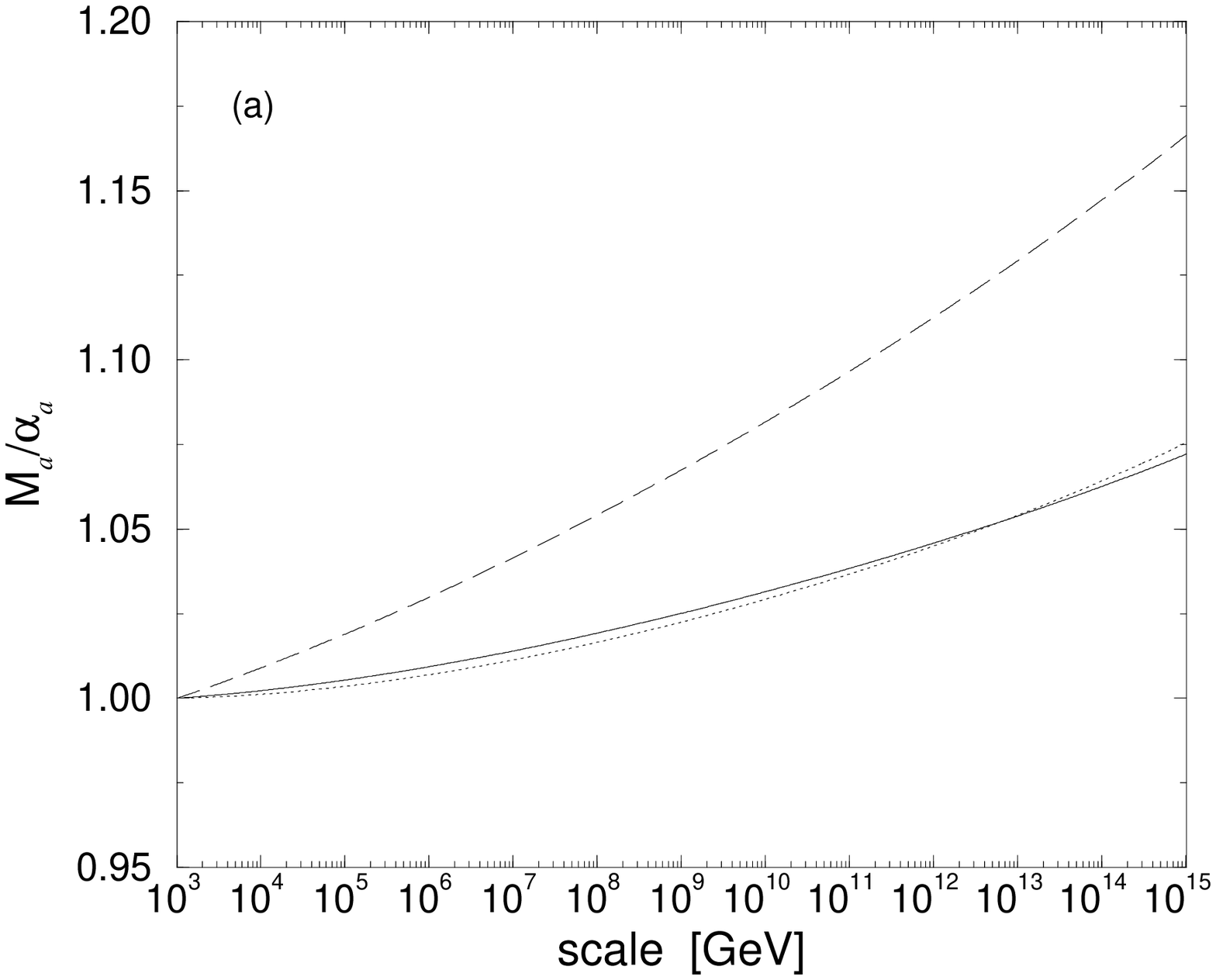}
\hfill
\epsfxsize=0.55\textwidth
\epsffile{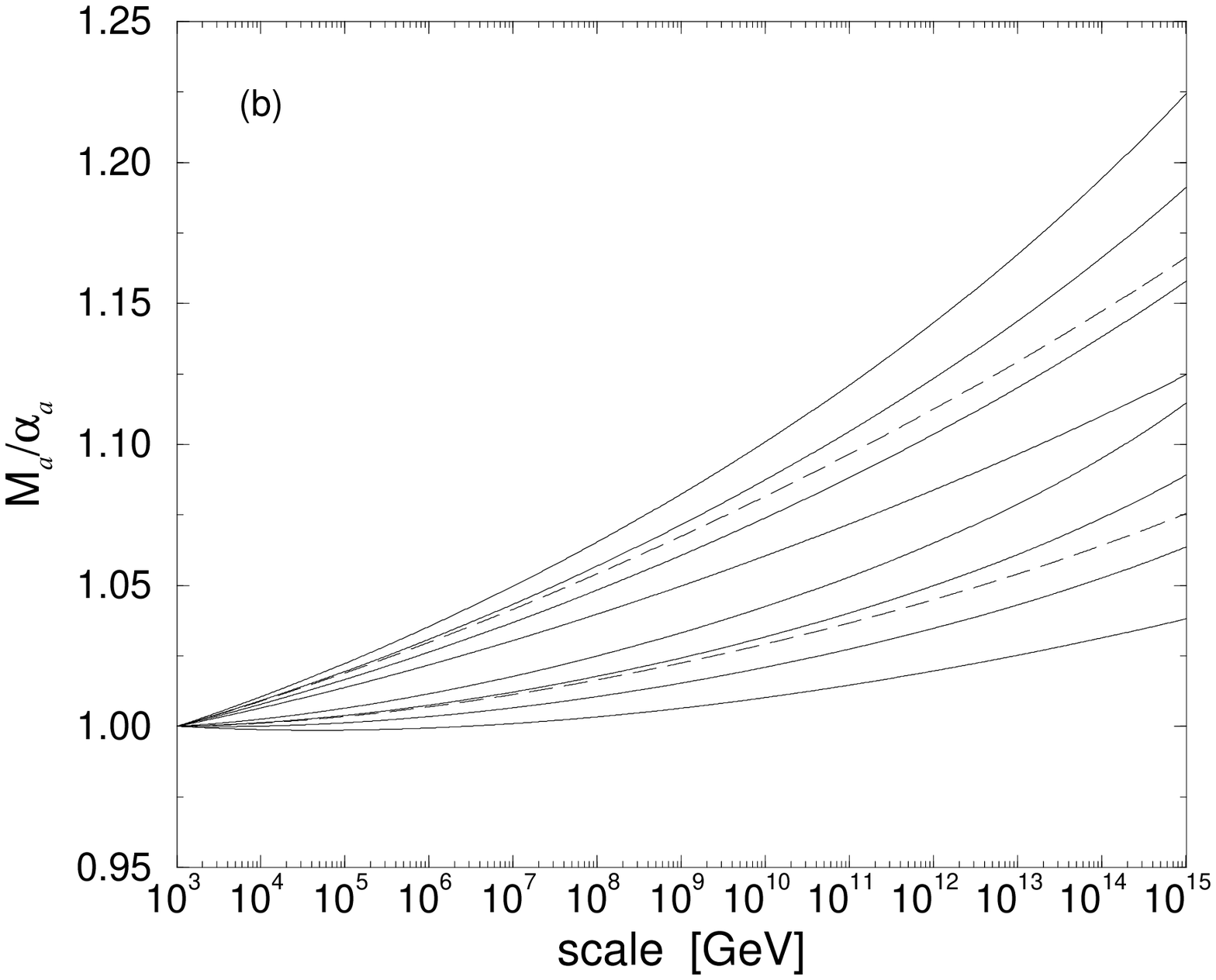}
\hfill }
\caption{Evolution of $\ratio{a}$ as a function of scale for
an $E_6$ model with (a) the Yukawa couplings $Y_S$ and $Y_D$
set to zero, and (b) the Yukawa couplings 
$Y_S(M_{\mathrm{unif}}) \approx Y_D(M_{\mathrm{unif}}) \approx 1$.  
In graph (a) the solid, dotted, and
dashed lines correspond to $a=1, 2, 3$.  In graph (b)
the four solid lines in the upper and lower part of the graph
correspond to $A_x = M_3, 0, -M_3, -2 M_3$ at $Q = 1$~TeV 
for $\ratio{3}$ and $\ratio{2}$ respectively.  For comparison
we also plotted $\ratio{3}$ and $\ratio{2}$ from Fig.~\ref{E6-fig}(a)
as dashed lines.  The ratios have been normalized against a
model without the extra matter needed to fill complete $\mathbf{27}$
representations of $E_6$.}
\label{E6-fig}
\end{figure}

There are four main effects that could alter the 
running of $\ratio{a}$ in this model: the extra matter, the
$U(1)_\eta$ gaugino mass, and the Yukawa couplings $Y_S$ and $Y_D$.
The simplest way to illustrate the effect of the
extra matter is to set the gauge coupling $g'$ and 
the Yukawa couplings $Y_S$ and $Y_D$ to be small.
Although some of the extra matter will obtain masses proportional
to these Yukawa couplings, we will simply work in the 
approximation where a common scale $Q_{\mathrm{extra}}$
can be chosen for the masses of the extra matter.
By this we mean the two generations of pairs of Higgs doublets 
that do not couple to quarks and leptons (the ``unHiggses''), 
all three generations of the color triplets, the right-handed
neutrinos, and the MSSM gauge singlets.  In Fig.~\ref{E6-fig}(a),
we show the evolution of $\ratio{a}$ in a model where,
again, the initial condition in Eq.~(\ref{init-cond-eq}) is
assumed.  The ratios $\ratio{a}$ are normalized against
an equivalent model with $Q_{\mathrm{extra}} > M_{\mathrm{unif}}$;
the latter is equivalent to a model without the $E_6$ matter content
(at the weak scale).  Since only the extra matter can alter the
evolution (when $g', Y_S, Y_D$ are small), Fig.~\ref{E6-fig}(a) 
shows that the three generations 
of color triplets induce the largest correction manifested in $\ratio{3}$,
although sizeable corrections are also present for $\ratio{1}$, $\ratio{2}$.
The gauge couplings $g_1$ and $g_2$ are nearly $30\%$ larger 
at\footnote{$M_{\mathrm{unif}}$ was somewhat arbitrarily 
chosen to be the where $g_1$ and $g_2$ meet, which is lower
in this $E_6$ model.}
$M_{\rm unif} \sim 3 \times 10^{15}$~GeV,
while $g_3$ is nearly $70\%$ larger,
as compared to a model without the extra matter.
In practice this means that the effects of the additional
$U(1)_\eta$ coupling $g'$ and the Yukawa couplings $Y_S$,
$Y_D$ are not nearly as significant to the RG evolution of $\ratio{a}$, 
since they compete against the considerably larger gauge couplings.
If we include the effects of the $U(1)_\eta$ gaugino,
the shift in the curves of Fig.~\ref{E6-fig}(a) is at the level
of about $0.007 c$, where $M' = c M_3$ at the weak scale.  Including Yukawa
couplings $Y_S(M_{\mathrm{unif}}) \approx Y_D(M_{\mathrm{unif}}) \approx 1$,
with scalar trilinear couplings in the range $-2 M_3 \le A_x \le M_3$,
there are more significant effects, as shown in 
Fig.~\ref{E6-fig}(b).  Again, the combined effects of large gauge 
couplings with larger Yukawa couplings push the scalar trilinear 
couplings to multi-TeV values at $M_{\mathrm{unif}}$.

Up to now we have tentatively placed the extra matter at 
$Q_{\mathrm{extra}} = 1$~TeV\@.  If the scale of the extra matter 
were increased to $2$, $10$~TeV, one finds the shift in $\ratio{a}$ 
at $Q = M_{\mathrm{unif}}$ is roughly $(-0.002, -0.002, -0.011)$, 
$(-0.019, -0.019, -0.046)$ for $a=(1,2,3)$ 
respectively, relative to the results in Fig.~\ref{E6-fig}(a).
Hence, increasing the scale of the extra matter by even a small 
amount (one order of magnitude of twelve), reduces the correction 
to $\ratio{a}$ by several percent.

\subsection{Extra matter}
\label{extra-matter-sec}
\indent

In the $\eta$-model of $E_6$ described above, it was necessary
to introduce extra matter to fill complete $\mathbf{27}$s
to ensure the cancellation of extra $U(1)$ anomalies.  Indeed, the 
extra matter dominated the correction to $\ratio{a}$.  However,
extra matter can, of course, be added without enlarging
the gauge structure of the model or even disrupting gauge
coupling unification, as long as anomaly cancellation with
the MSSM group is ensured.  
A recent, widely touted example of adding extra matter without
upsetting gauge-coupling unification is
the addition of $\mathbf{5} + \overline{\mathbf{5}}$ pairs 
(or a $\mathbf{10} + \overline{\mathbf{10}}$ pair) used as the 
messenger sector of gauge-mediated models~\cite{Dine, GRreport}.
While we defer going into details about gauge-mediation 
(particularly the supersymmetry breaking masses associated
with the messenger matter) until Sec.~\ref{gauge-mediation-sec},
this does serve as one interesting starting point for
adding extra matter.  Nevertheless, for the remainder of the section
the extra matter is assumed to have only supersymmetry preserving
masses.  

It is perhaps useful to review a few recent examples of the uses of
extra matter for model-building and phenomenology
(other than gauge-mediation).
Ref.~\cite{MartinRamond} showed that adding extra matter 
could ``refocus'' the gauge couplings to unify at a higher scale.
The motivation was an attempt to bridge the discrepancy between
$M_{\mathrm{unif}}$ and $M_{\mathrm{string}}$, and indeed
they were able to show that only particular kinds of extra
matter were able to do the job.  In Ref.~\cite{KoldaMR} a 
``semi-perturbative'' model was constructed whereby 
the gauge couplings became large, but still perturbatively
calculable, at the high scale.  They added sufficient
$\mathbf{5} + \overline{\mathbf{5}}$ and 
$\mathbf{10} + \overline{\mathbf{10}}$ pairs to nearly
saturate the $\beta$-functions, and found the interesting
result that the gaugino mass ratios $M_a/M_b$ at the weak
scale could be quite different from the canonical
expectations of models without the extra matter.
By adding matter in complete representations of $SU(5)$,
gauge-coupling unification (to one-loop) is not upset.
However, as pointed out in Ref.~\cite{Martin},
extra matter can be added in other combinations that also
maintain (one-loop) gauge coupling unification, 
indicating that there are a variety of possibilities 
of extra matter that could be explored.  Finally,
Ref.~\cite{GLR} showed that adding extra matter
in complete $\mathbf{5} + \overline{\mathbf{5}}$ and 
$\mathbf{10} + \overline{\mathbf{10}}$ does not 
change\footnote{With up to intermediate values of the
gauge couplings at the unification scale.  For much larger values
the expectations are quite different~\cite{KoldaMR, GLR2}.}
the $\alpha_s$ value extracted at the weak-scale,
but could increase the unification scale.  In particular, 
by taking the heavy multiplets to have a common mass at the 
high scale but split in mass at intermediate scales (by RG gauge 
corrections), they showed that 
the one-loop threshold corrections largely cancel 
the two-loop corrections in the $\beta$-functions.
Although this is important to determine (separately) the gauge 
couplings or gaugino masses at the high scale, the one-loop 
threshold corrections cancel in the ratio $M/\alpha$.
Thus, in the examples to follow, we will generally assume the 
extra matter is introduced with a common mass at a single 
intermediate scale.

Extra matter can be safely added in vectorlike pairs,
allowing an arbitrarily large supersymmetric mass term
and ensuring the cancellation of anomalies.  We therefore
have two sets of parameters; the quantity (and type) of
extra matter, and the scale(s) where it is introduced.
Using the results in Appendix~\ref{RGE-app}, extra matter
can be easily incorporated by merely shifting
the $\beta$-function coefficients at the scale of the
new matter, creating a new effective theory.  Increasing the 
amount of extra matter can be compensated by increasing the 
scale where the matter is introduced, and so following 
Ref.~\cite{KoldaMR} we introduce an \emph{effective}
amount of extra matter $n_{\mathrm{eff}}$ that (in particular)
need not be an integer.  In this way, extra matter need only be added 
at one scale (that we take to be near the weak scale $\sim 1$~TeV),
and then the size of the deviation in $\ratio{a}$ is computed
as a function of the continuous parameter $n_{\mathrm{eff}}$.

The type of extra matter introduced could be of many different
varieties, however there are constraints.  If we restrict ourselves 
to vectorlike matter, then there could be vectorlike pairs
of MSSM matter (e.g. $Q + \overline{Q}$), or limitless varieties
of exotics that may (or may not) fill representations of 
common GUT groups, and that may (or may not) preserve gauge 
coupling unification.  We will consider
only four scenarios of adding extra matter, and we decided
to examine scenarios that do not upset gauge coupling unification
(at least to one-loop).  The scenarios are:
\begin{eqnarray}
\mathrm{Set \; 1:} & & n_{\mathrm{eff}} \; = \; n_{\mathbf{5}} 
   + n_{\mathbf{\overline{5}}} \\
\mathrm{Set \; 2:} & & n_{\mathrm{eff}} \; = \; n_{\mathbf{10}} 
   + n_{\mathbf{\overline{10}}} \\
\mathrm{Set \; 3:} & & n_{\mathrm{eff}} \; = \; 2 n_Q \; = \; 2 n_d 
   \; = \; 4 n_e \\
\mathrm{Set \; 4:} & & n_{\mathrm{eff}} \; = \; 2 n_Q \; = \; 4 n_u 
   \; = \; 2 n_L \; .
\end{eqnarray}
Set~1 and Set~2 are the common extensions of adding vectorlike
matter in complete $SU(5)$ representations
$\mathbf{5} + \mathbf{\overline{5}}$ and 
$\mathbf{10} + \mathbf{\overline{10}}$ pairs, respectively.
We also consider adding
\begin{eqnarray}
\mathrm{Set \; 3:} & & (\mathbf{3}, \mathbf{2}, \textfrac{1}{6}) + 
(\mathbf{\overline{3}}, \mathbf{1}, \textfrac{1}{3}) +
2 \times (\mathbf{1}, \mathbf{1}, 1) + conj.
\end{eqnarray}
and 
\begin{eqnarray}
\mathrm{Set \; 4:} & & (\mathbf{3}, \mathbf{2}, \textfrac{1}{6}) + 
2 \times (\mathbf{\overline{3}}, \mathbf{1}, -\textfrac{2}{3}) +
(\mathbf{1}, \mathbf{2}, -\textfrac{1}{2}) + conj.
\end{eqnarray}
which do not apparently form complete representations of
any simple GUT group (Set~3 was first considered in 
Ref.~\cite{Martin}).  Nevertheless, Sets~3 and 4 ensure
the one-loop $\beta$-functions are shifted by the same amount
independent of the particular coupling, and therefore one-loop
gauge coupling unification is preserved.

\begin{figure}[!t]
\centerline{
\hfill
\epsfxsize=0.55\textwidth
\epsffile{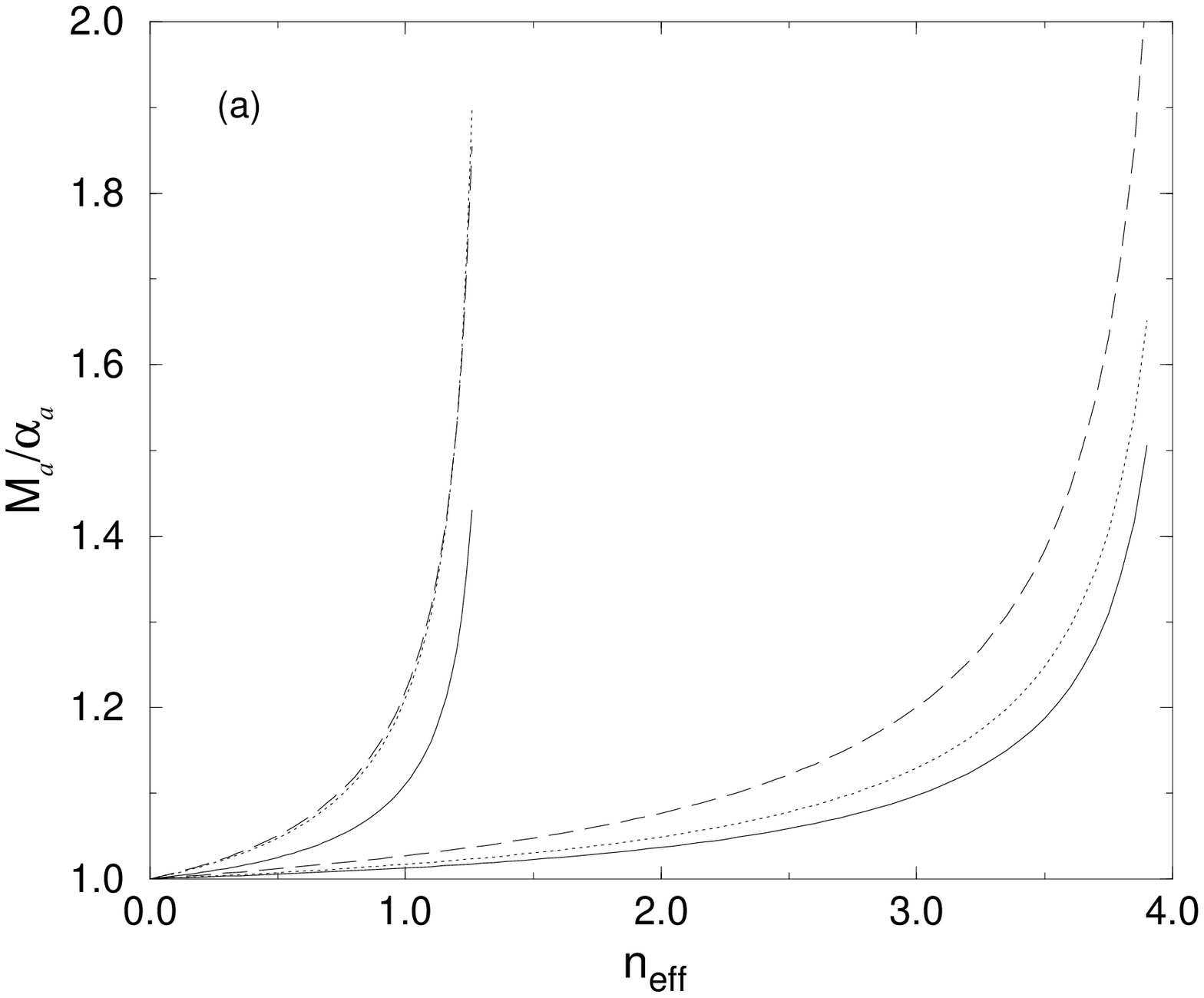}
\hfill
\epsfxsize=0.55\textwidth
\epsffile{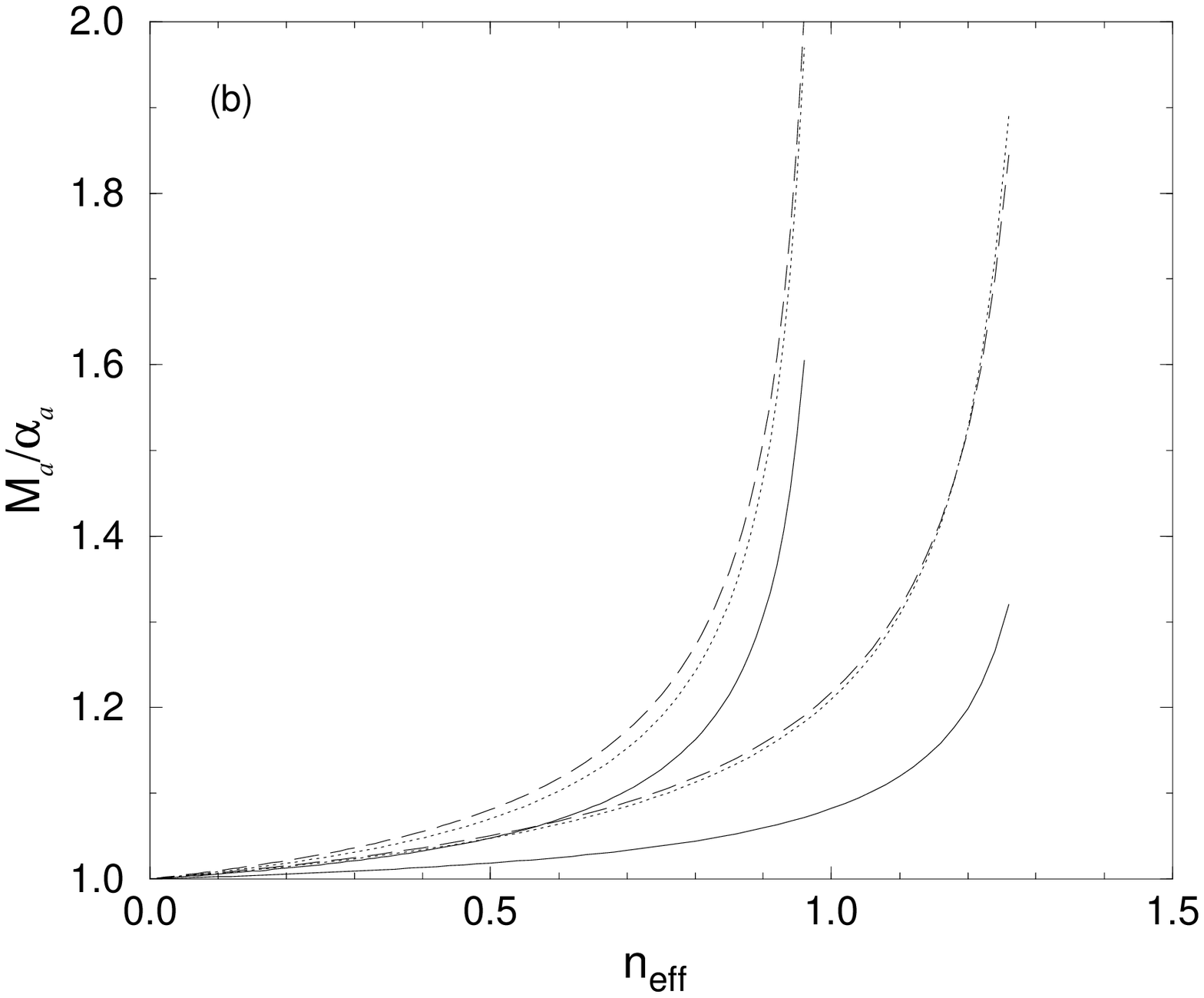}
\hfill }
\caption{$\ratio{a}$ evaluated at $Q = 10^{16}$~GeV with four classes
of extra matter, normalized to a model without extra matter.
The solid, dotted, and dashed lines correspond to $a=1,2,3$.
Graph (a) contains $n_{\mathrm{eff}}$ number
of $\mathbf{5} + \mathbf{\overline{5}}$ (triplet of lines on the
right) and $n_{\mathrm{eff}}$ number of $\mathbf{10} + 
\mathbf{\overline{10}}$ (triplet of lines on the left), and 
graph (b) contains $n_{\mathrm{eff}}$ number 
of Set~3 (triplet of lines on the right) and $n_{\mathrm{eff}}$ 
number of Set~4 (triplet of lines on the left).}
\label{effective-fig}
\end{figure}

In Fig.~\ref{effective-fig} we show the effect of adding
$n_{\mathrm{eff}}$ number of pairs of the additional matter in
one of the above scenarios.
The curves are cut off when any coupling $\alpha$ exceeds the 
somewhat arbitrary value $0.5$; beyond this value, perturbative
calculations are probably not reliable~\cite{KoldaMR}.
It is clear that the size of the correction to $\ratio{a}$
by adding extra matter can be very large, in comparison to all the
previous effects that have been considered.  However, this occurs
only when the gauge couplings are near their semi-perturbative
value, and indeed we were unable to find corrections much larger
than $20\%$ to $30\%$ without a gauge coupling exceeding
$\alpha \sim 0.2$ at $10^{16}$~GeV\@.  In all cases
above, the largest correction occurred for $\ratio{3}$
due to our choice of adding extra matter that preserves 
one-loop gauge coupling unification, that implies $g_3$ remains
the largest gauge coupling throughout the RG evolution.

The effect of increasing the mass $M_{\mathrm{extra}}$ of the 
extra matter can be easily approximated by rescaling 
$n_{\mathrm{eff}}$ by a factor 
\begin{eqnarray}
n_{\mathrm{eff}} &\ra& \frac{\ln \frac{Q_{\mathrm{high}}}{1 \; 
   \mathrm{TeV}}}{\ln \frac{Q_{\mathrm{high}}}{M_{\mathrm{extra}}}} 
   n_{\mathrm{eff}} \; ,
\end{eqnarray}
where $Q_{\mathrm{high}} \sim 10^{16}$~GeV\@.  If the amount
of running is reduced $Q_{\mathrm{high}} < 10^{16}$~GeV, 
the rescaling of $n_{\mathrm{eff}}$ is more complicated because
the three gauge couplings are no longer expected to be equal 
at the scale where the running is stopped.

\section{High scale corrections}
\label{high-scale-sec}
\indent

There are a variety of effects that can upset the ratio $\ratio{a}$
near the high scale.  In the following, we will examine the
well-known effects of GUT scale threshold corrections~\cite{HMG}, 
discuss issues related to the scale of GUT breaking,
examine Planck scale corrections, and finally discuss effects 
unique to supergravity.

\subsection{GUT scale thresholds}
\label{GUT-thresholds-sec}
\indent

If gauge coupling unification is indeed indicative of a
grand unified theory, then many additional heavy fields near
the high scale will be present.  In $SU(5)$, for example, 
there are the heavy gauge boson and gaugino remnants
of the adjoint, transforming as $(\mathbf{3}, \mathbf{2}, 
\textfrac{5}{6})$ under $SU(3)_c \times SU(2)_L \times U(1)_Y$, 
as well as heavy Higgs multiplets (from, e.g.\ the $\mathbf{24}$
and $\mathbf{5} + \overline{\mathbf{5}}$ of Higgs), and
possibly heavy chiral multiplets as well.  The corrections
to the ratio $\ratio{a}$ due to integrating out the heavy
remnants of GUT breaking were calculated in Ref.~\cite{HMG},
and we will follow their discussion closely.
There are two classes of corrections to the ratio $\ratio{a}$: 
wave function renormalization and mass renormalization.
For a heavy chiral or vector supermultiplet with a mass
$M_{\mathrm{heavy}}$, the wave function renormalization corrections
give rise to logarithmic corrections to $M_a$ and $g_a$
that scale as
\begin{equation}
\sim \; \frac{\alpha}{4 \pi} 
     \ln \frac{M_{\mathrm{heavy}}^2}{Q_{\mathrm{GUT}}^2} \; ,
\end{equation}
after integrating out the heavy fields
near the GUT scale.  Of course this is completely 
equivalent to modifying the $\beta$-function coefficients at
the scale of the heavy fields, and therefore the logarithmic
corrections cancel in the ratio $\ratio{a}$~\cite{HMG, KoldaMR}
to one-loop.  (Two-loop corrections can be neglected since
we are assuming the heavy fields are near the GUT scale,
and so no large logarithms are present.)

The second class of corrections arise when soft supersymmetry 
breaking masses are introduced in the mass matrices for
the heavy fields, leading to mass renormalization.
In a general supersymmetric theory
with soft breaking, we would expect soft terms for all
sparticles with a scale of order the breaking scale
$M_{\mathrm{SUSY}}$, presumably near the weak scale.  
The soft breaking masses are ordinarily insignificant compared 
with the supersymmetry preserving masses induced from GUT 
breaking.  However, the corrections to $\ratio{a}$ take the 
form~\cite{HMGnote}
\begin{eqnarray}
\frac{M_a}{\alpha_a} 
   &=& \frac{M_5(Q_{\mathrm{GUT}})}{\alpha_5(Q_{\mathrm{GUT}})} 
   + \frac{1}{4\pi} \left( 2 C(G_a) [M_5(Q_{\mathrm{GUT}}) - \delta m] 
   + \sum_R S_a(R) B_R \right)
\end{eqnarray}
where $M_5(Q_{\mathrm{GUT}})$ and 
$\alpha_5(Q_{\mathrm{GUT}})$ are the unified 
gaugino mass and gauge coupling at the GUT scale $Q_{\mathrm{GUT}}$, 
$\delta m$ is the mass of the fermion component of the Nambu-Goldstone
multiplet induced by supersymmetry breaking, and $B_R$ are
the standard $B$-terms in the soft supersymmetry breaking
Lagrangian for the chiral representations $R$.
The group theory factors are the quadratic Casimir 
$C(G_a)$ for the adjoint $G_a$, and the Dynkin index $S_a(R)$ 
for the chiral representation $R$, both for the group
associated with the ``$a$'' gaugino.  When the supersymmetry 
breaking masses $M_5$, $\delta m$ and $B_R$ are roughly the same 
scale as the gaugino masses, the correction is typically of 
order a few percent.

In the case where the GUT scale chiral multiplets fill 
up a complete representation of the GUT group, the
corrections $\sum_a S_a(R) B_R$ are finite but universal.
This is completely analogous to integrating out the messenger 
sector of gauge-mediated models, and indeed the calculation
of the gaugino mass in gauge-mediated models~\cite{Dine, Martin}
is identical to the calculation above, in the limit that the 
mass splitting of the scalar components of the messenger 
superfields is much smaller than the associated fermion mass.
The presence of heavy fields with supersymmetry breaking masses
is a bona fide possibility in itself, and the heavy fields 
need not be associated with a GUT nor do they need to 
have masses near the GUT scale (although a GUT does
provide a strong motivation for the existence of such fields).
Thus GUT scale corrections to $\ratio{a}$ are indeed 
possible, but the size of the effect depends solely on the 
size of the supersymmetry breaking masses associated 
with the heavy fields.

\subsection{Breaking GUTs at other scales}
\label{GUT-breaking}
\indent

When we discussed the $E_6$ GUT in Sec.~\ref{eta-E6-sec}, it was 
assumed that the breaking pattern $E_6 \ra SU(5) \times U(1) 
\times U(1) \ra SU(3)_c \times SU(2)_L \times U(1)_Y \times U(1)_\eta$
occurred near the high scale.  Extrapolating $\ratio{a}$ 
(for $1 \le a \le 4$) was therefore well-defined up to 
this GUT breaking scale.  However, it is very possible that the 
breaking of the GUT group does not go directly to the standard 
model (plus extra $U(1)$'s), but instead occurs in stages.  For example,
the rank-6 group $E_6$ could break to either $SO(10) \times U(1)$
or $[SU(3)]^3$, the latter breaking to 
$SU(3)_c \times SU(2)_L \times SU(2)_R \times U(1)_{B-L}$
and then to the MSSM group.  The rank-5 group $SO(10)$ could break 
to $SU(5) \times U(1)$ or $SU(4)_c \times SU(2)_L \times SU(2)_R$, 
and then there are various permutations of breaking the latter
$SU(4)_c$ and $SU(2)_R$.
The phenomenology of these breaking patterns has been well-studied
(for a review, see Ref.~\cite{Mohapatra}), and for our 
purposes it is important to recognize that the ratios $\ratio{a}$
may only have meaning up to some intermediate scale, where part or 
all of the MSSM group is embedded in a more complicated
(possibly higher rank) structure.  It was pointed out in
Ref.~\cite{KMY} that despite the possibly complicated
breaking patterns in $SO(10)$, the expectations for the gaugino 
mass ratios at the weak scale to one-loop were identical
to the case where the breaking is directly to the MSSM group.  
Two-loop corrections are probably small, unless the gauge
or Yukawa couplings become large, or unless the two-loop
coefficients are significantly altered by the extra 
matter in higher dimensional representations needed to accomplish 
the various stages of breaking.  Above the breaking scale, 
the ratios $\ratio{a}$ are subsumed into ``new'' one-loop 
invariants $M/\alpha$ associated with the group at that
scale [e.g.\ $\ratio{5}$ for $SU(5)$ or $\ratio{10}$ for $SO(10)$].  
Although we have not attempted to analyze the various
(two-loop and other) effects that could disrupt these GUT
one-loop invariants, there would be nonzero corrections.

\subsection{Planck scale corrections}
\indent

It is well known that physics near the Planck scale can affect
low energy predictions of, for example, the gauge couplings
through higher dimension operators~\cite{Hill, Shafi}.  
These analyses presume quantum gravity induces nonrenormalizable 
(dimension $> 4$) terms suppressed by the Planck 
scale\footnote{Normally taken to be the ``reduced'' Planck 
scale $M = M_{Pl} / \sqrt{8\pi} \approx 2.4 \times 10^{18}$~GeV.}, 
that can take the form~\cite{Hill}
\begin{equation}
\frac{1}{M} \left[ \tr (W^\alpha W_\alpha \Sigma) + h.c. \right]_F \; ,
\label{Planck-operator-eq}
\end{equation}
where $W^\alpha$ and $\Sigma$ are gauge and Higgs superfields
and the trace is over the gauge group generators.
If $\Sigma$ acquires a vev for its scalar component $S$
(through, e.g., the breaking of a GUT gauge group),
then the kinetic terms for the gauge fields are modified
\begin{eqnarray}
\delta \mathcal{L}_{\mathrm{kinetic}} &=& 
    - \frac{1}{4 M} \tr (F_{\mu\nu} F^{\mu\nu} S ) \; ,
\end{eqnarray}
where $F_{\mu\nu} \equiv {F_{\mu\nu}}_a T^a$ is the
field strength tensor of the unified gauge group.
To bring the kinetic terms back into a canonical form,
the fields are rescaled, and therefore the gauge couplings
must also be rescaled.  This is the well-known Planck scale
correction to the gauge couplings, that has been discussed
in Refs.~\cite{Hill, Shafi, LP1, HallSarid, Arnowitt, Nath}.

The Planck scale operator not only gives modifications to the 
gauge kinetic terms, but also modifies the gaugino kinetic 
terms (and auxiliary fields)~\cite{Hill, EENT, DreesSUGRA}
\begin{eqnarray}
\delta \mathcal{L}_{\mathrm{kinetic}} &=&
    \frac{1}{M} \tr ( i \lambda \slashchar{\mathcal{D}} 
    \overline{\lambda} S + \textfrac{1}{2} D^2 S ) \; ,
\end{eqnarray}
where $\lambda$ is the 2-component gaugino field,
$D$ is the auxiliary field, and $\mathcal{D}_\mu$ is the 
covariant derivative.  To bring these terms back into
a canonical form, the gaugino (and auxiliary) fields must also
be rescaled.  Hence, the soft term
$\textfrac{1}{2} M (\lambda\lambda + h.c.)$ is
rescaled, giving a correction to the gaugino mass simultaneously 
with the correction to the gauge coupling.  It is
simple to show that the correction to $M$ cancels in the
ratio with $g^2$, and therefore Planck scale operators
of the form in Eq.~(\ref{Planck-operator-eq}) do not affect 
the ratio $M/\alpha$.  This is easily generalized to $\ratio{a}$.
Furthermore, any rescaling of the 
gauge kinetic terms (from higher dimension operators, or any
other source) must be accompanied by rescaling of the 
gaugino kinetic terms by the same amount and therefore cancels
in the ratio $M/\alpha$, if supersymmetry is to be preserved.

\subsection{Supergravity effects}
\indent

In the previous section concerning Planck scale corrections
of the type in Eq.~(\ref{Planck-operator-eq}), the operator 
was implicitly assumed to be \emph{globally} supersymmetric.  
When this is generalized to a
locally supersymmetric theory (supergravity), the Lagrangian becomes 
more complicated~\cite{Cremmer}.  Specifically,
the gauge kinetic terms plus gaugino soft terms take the form
\begin{eqnarray}
\mathcal{L} \subset 
    -\textfrac{1}{4} \re f_{ab} F^a_{\mu\nu} F^{\mu\nu b}
    -\textfrac{1}{4} \re f_{ab} \overline{\Lambda}^a 
         \slashchar{\mathcal{D}} \Lambda^b
    +\textfrac{1}{4} e^{-G/2} G^i (G^{-1})_i^j 
         \frac{\partial f_{ab}^*}{\partial {\phi^j}^*} 
    \overline{\Lambda}^a \Lambda^b
\end{eqnarray}
where two new functions are introduced; the gauge kinetic
function $f_{ab}$, and the K\"{a}hler potential $G$, 
with derivatives denoted by $G^i \equiv \partial G/\partial \phi_i$
and $G_i^j \equiv \partial^2 G /\partial {\phi^*}^i \partial \phi_j$.
($\Lambda$ is the 4-component Majorana gaugino field and $\phi_i$
is the scalar component of the chiral superfields $\Phi_i$ 
in the theory.)
The gauge kinetic function $f_{ab}$ can be thought of 
as parameterizing the dimension $> 4$ Planck scale operators.
The K\"{a}hler potential is indigenous to supergravity,
and is a (real) function of the scalar components of the chiral
superfields in the theory.
Since its form is \emph{a priori} unknown, the coefficient
of the gaugino soft mass term is unknown.  Scalar fields 
with vevs near the Planck scale are expected to modify 
the K\"{a}hler potential, and therefore modify the gaugino 
masses \emph{independent} of the kinetic terms~\cite{EENT, 
DreesSUGRA, Nath}.  
Thus, the relation $M_a/\alpha_a$ is expected to be modified 
when embedded in a supergravity theory with a non-minimal 
K\"{a}hler potential.  This should not be surprising, since
universal nonzero gaugino masses are accomplished by specifying
$f_{ab} = S \delta_{ab}$ and $G_i^j = -\delta_i^j$ (a flat 
K\"{a}hler manifold), and determined by the gravitino
mass proportional to $\exp(-\langle G \rangle/2)$.
More detailed computations of the gaugino masses
in stringy supergravity models were carried out in 
Ref.~\cite{string-model-gaugino-mass}
to which we refer the interested reader.

\section{Gauge-mediated masses}
\label{gauge-mediation-sec}
\indent

The final class of ``corrections'' to $\ratio{a}$ that we will 
consider is from extra matter with supersymmetry breaking masses.
Ordinarily supersymmetry breaking is assumed to occur in a
``hidden'' sector, which is assumed to contain matter
with no MSSM gauge charges.  Consequently, only gravitational 
interactions exist between the hidden sector and the MSSM, and
it is through gravity that supersymmetry breaking is communicated
to the ``visible'' sector; this is the canonical 
``supergravity-inspired'' model.
An alternative possibility is that supersymmetry is broken in 
a sector that does have couplings to the MSSM, 
and thus ordinary gauge interactions can
communicate supersymmetry breaking to the MSSM\@.  This
scenario has been the study of an enormous amount of recent 
work~\cite{Dine, GRreport}.

In gauge-mediated models, a ``messenger'' sector is introduced that has
both supersymmetry preserving masses and supersymmetry breaking 
masses.  The supersymmetry preserving masses ensure that the messenger 
masses are considerably larger than the weak scale, while the 
supersymmetry breaking masses can successfully induce weak scale 
masses in the MSSM sparticles.  The canonical approach~\cite{Dine} 
is to assume the messengers form complete representations of $SU(5)$ to
ensure gauge coupling unification is not disrupted at one-loop
(however, see Ref.~\cite{Martin}).
The superpotential for the messenger sector takes the form
\begin{eqnarray}
W &=& \lambda S \Phi \overline{\Phi}
\label{mess-superpotential-eq}
\end{eqnarray}
where $\Phi$, $\overline{\Phi}$ transform as a vectorlike 
representation of the MSSM gauge group, such as
\begin{equation}
(\mathbf{3}, \mathbf{1}, -\textfrac{1}{3}) + 
(\mathbf{1}, \mathbf{2}, \textfrac{1}{2}) + conj.
\end{equation}
that make up the $\mathbf{5} + \mathbf{\overline{5}}$ of $SU(5)$.  
Supersymmetry 
breaking occurs when the auxiliary component of the gauge singlet 
$S$ acquires a vev $F_S$, but we also assume the scalar component 
of $S$ also acquires a vev $\langle S \rangle$
to give $\Phi$ and $\overline{\Phi}$ (large) 
supersymmetry preserving masses.  The spin-1/2 components of
$\Phi$ and $\overline{\Phi}$ are combined as one single
Dirac fermion, with a mass $m_f \equiv \lambda \langle S \rangle$.
The two complex scalars from $\Phi$ and $\overline{\Phi}$ 
acquire masses $m_1$ and $m_2$,
with $m_{1, 2}^2 = m_f^2 \mp \lambda F_S$.  In general, additional 
(higher dimensional) interactions could be added to 
Eq.~(\ref{mess-superpotential-eq}); the effect of such terms is to 
modify the relation between $m_{1, 2}$ and $m_f$, parameterized
by $\mathrm{Str} \; M^2_{\mathrm{mess}} \equiv 2 m_1^2 + 2 m_2^2 - 
4 m_f^2$~\cite{Poppitz}.  

\begin{figure}
\begin{picture}(457,45)(0,0)
  \Line( 160, 10 )( 300, 10 )
  \Photon( 160, 10 )( 200, 10 ){3}{4}
  \Photon( 260, 10 )( 300, 10 ){3}{4}
  \DashCArc( 230, 10 )( 30, 0, 180 ){3}
  \Text( 230, 10 )[c]{\Large $\times$}
  \Text( 230, 40 )[c]{\Large $\times$}
\end{picture}
\caption{The one-loop gaugino mass induced by heavy messengers.
The crosses represent the relevant mass insertions.}
\label{messenger-fig}
\end{figure}
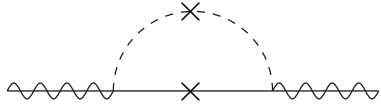

The one-loop messenger contributions to the gaugino masses have been
presented in Refs.~\cite{Dine, Martin}, and generalized to
the case $\mathrm{Str} \; M^2_{\mathrm{mess}} \not= 0$ 
in Ref.~\cite{Poppitz}.
The relevant diagram is shown in Fig.~\ref{messenger-fig},
which is equivalent to the mass renormalization diagram due to
heavy chiral multiplets discussed in Sec.~\ref{GUT-thresholds-sec}, 
and is the supersymmetry breaking analog to Fig.~\ref{gaugino-fig}(a).
The result for the self-energy can be written as
\begin{eqnarray}
\Sigma(\slashchar{p}) &=& \frac{\alpha}{4 \pi} m_f S(\Phi) \sin 2\theta
    \times \left[ \frac{y_2}{1 - y_2} \ln y_2 - \frac{y_1}{1 - y_1} \ln y_1 
    \right]
\label{gaugino-sigma-eq}
\end{eqnarray}
where $y_i = m_i^2/(p^2 + m_f^2)$, $m_f$ is the messenger fermion mass,
$m_1$ and $m_2$ are the messenger scalar masses, $\theta$ is
the angle that diagonalizes the mass matrix\footnote{We assume
the scalar mass matrix is real.}, and $S(\Phi)$ is the usual Dynkin
index for the messenger superfield $\Phi$.  The correction to
the gaugino mass can be approximated as simply
\begin{eqnarray}
M &=& M_0 + \Sigma(M_0) 
\label{gaugino-mass-eq}
\end{eqnarray}
where $M$ and $M_0$ is the gaugino mass with and without the
supersymmetry breaking sectors integrated out.
The expression given in Eqs.~(\ref{gaugino-sigma-eq}), 
(\ref{gaugino-mass-eq}) is a trivial generalization of the 
formulae given in Refs.~\cite{Martin, Poppitz}, and reduces
to the latter formulae in the limit that the gaugino mass is induced 
entirely from messenger interactions.  In fact, the $p^2$ correction
in the denominator of $y_i$ is negligible when the messenger fermion 
mass is much larger than the gaugino mass, which is the usual
situation.

The size of the correction to the gaugino mass is dependent
entirely on the size of the supersymmetry breaking mass $\sqrt{F_S}$
in the messenger sector.  In the case of heavy chiral multiplets
with supersymmetry breaking masses of order the weak scale~\cite{HMG},
the correction is of order a few percent.  This is because
the one-loop self-energy correction $\Sigma(M_0)$ is suppressed
by $\alpha/(4\pi)$.  In gauge-mediated
models, this suppression is overcome by specifying a supersymmetry
breaking mass of order the weak scale times $4 \pi/\alpha$,
which is of order tens of TeV\@.  Note that two-loop corrections 
to the gaugino masses have also been recently computed in 
Ref.~\cite{Strumia}, and can give corrections up to $10\%$
to the one-loop result if the messenger scale is of order
the supersymmetry breaking mass.

It is also worth considering the possibility that extra heavy
vectorlike matter exists with supersymmetry breaking 
masses\footnote{To avoid the introduction of additional
complications, the supersymmetry breaking can be assumed to
be introduced explicitly.  Those who find this unpalatable
can imagine gauge singlet interactions (plus possibly higher
dimensional terms) analogous to Eq.~(\ref{mess-superpotential-eq}).}
that do not come in complete representations of $SU(5)$, nor
even preserve gauge-coupling unification.  This class of 
matter would communicate its supersymmetry breaking to 
only those sparticles charged under the appropriate groups
to which the extra matter transforms nontrivially.  There are, however, 
some important comments.  The size of the contribution to any 
given gaugino mass (and thus the ratio $\ratio{a}$) is 
determined by the size of the supersymmetry breaking
mass (and the multiplicity of the extra matter).
The correction to the gauge couplings (at the high scale)
due to the extra matter is determined solely by the multiplicity and
the scale of the extra matter, i.e.\ the supersymmetry 
preserving masses\footnote{Highly split supermultiplets
will have additional corrections dependent on the
size of the supersymmetry breaking masses.}.  Thus,
some classes (or multiplicities) of extra matter could be 
excluded if the gauge couplings are required to be perturbative
to the high scale.  The size of the correction to the gaugino
mass is, however, undetermined.  Hence, the correction 
to $\ratio{a}$ could be arbitrarily large.

There are, of course, phenomenological consequences if
supersymmetry is broken at scales smaller than the
canonical scale $\sim 10^{11}$~GeV assumed in supergravity models.
In particular, a light gravitino is expected.
The presence or absence of a light gravitino in
collider experiments would help determine (or at least
put a lower bound on) the fundamental 
scale of supersymmetry breaking (see e.g.\ \cite{DDRT, AKKMMgrav}).
However, it probably does not determine the supersymmetry 
breaking masses in the extra (messenger) matter because generically 
one expects complicated feed down mechanisms in the dynamical 
breaking sectors that ultimately separate the ``true'' 
scale $\sqrt{F}$ from $\sqrt{F_S}$~\cite{Dine}.

\section{Conclusions}
\label{conclusions-sec}
\indent

In this paper, we have studied the effects that can disrupt
the one-loop renormalization group invariant, $\ratio{a}$.
This includes two-loop corrections in the MSSM, weak scale
threshold corrections, and superoblique corrections.  Two-loop 
corrections from extensions of the MSSM within a minimal $U(1)$ model,
an $E_6$ model, and extra matter were also considered.  In addition, 
high scale corrections from GUT thresholds, Planck scale operators, 
and supergravity effects were discussed.  Finally, the
effects from a heavy sector with supersymmetry breaking masses
that communicates via gauge-mediation to the MSSM were 
discussed.  

The two-loop corrections (both within the MSSM, and beyond)
were, not surprisingly, found to be large only when a coupling becomes 
large near the high scale.  Otherwise, the corrections are of
order a few to tens of percent after evolving $13$ orders
of magnitude in scale.  The ``pure gaugino'' terms were
usually small except for the $g_3^2 M_3$ term, and in particular
the extra $U(1)'$ gaugino in extended models typically gives a very
small correction.  The Yukawa terms can induce a larger correction
(up to ten to perhaps tens of percent at the high scale), particularly
if $\tan\beta$ is near one of its perturbative extremes.
Typically, to obtain a larger correction the scalar trilinear 
couplings must run to several times their weak scale values
at the high scale.
The Yukawa terms, and hence the correction to $\ratio{a}$, 
can take either sign depending on the sign of the scalar 
trilinear coupling (although note that the sign of the
scalar trilinear coupling is \emph{not} RG invariant).
The corrections from extra matter can also be substantial,
becoming very large (factor of $1.5$ to $2$) if the 
gauge couplings approach their semi-perturbative values.  

We also showed that superoblique corrections cancel in
the ratio $M_a/{\tilde{g}}_a^2$, where $\tilde{g}_a$ is
the coupling of the gaugino to the scalar and fermionic
piece of a chiral superfield (that transforms nontrivially
under the associated gauge group).  However, since experiments
measure the gauge couplings extracted from gauge boson
interactions (at least for the foreseeable future), 
there can be a significant correction
in translating $\ratio{a}$ into the true one-loop invariant
$M_a/{\tilde{g}}_a^2$, if some supermultiplets are widely 
split in mass. 

Weak scale threshold corrections proportional to a 
(possibly large) logarithm cancel to one-loop in the ratio 
$\ratio{a}$.  There are, however, residual finite effects when 
translating measured pole masses into $\overline{\mathrm{DR}}$ 
running masses, and effects from gaugino mixing. 
Thus, complete weak scale threshold corrections would be 
necessary for high precision.  To compute the size of
other effects considered in this paper, we specified the 
\emph{running} masses at a common scale ($1$~TeV) above weak
scale thresholds.  This was mainly
for computational convenience since the complete weak scale threshold
corrections can become rather complicated; of course one can always 
do the matching of our initial conditions to a particular weak 
scale model if desired.

High scale effects were also discussed.  We reviewed GUT threshold 
corrections that have been calculated before, noting
that the correction depends only on the soft breaking 
masses of the heavy GUT fields.  GUT breaking that occurs
in stages (or at a smaller scale) was also mentioned, implying 
the ratios $\ratio{a}$ may only be applicable up to a scale that
could be considerably smaller than the unification scale.  Planck scale 
corrections from one commonly considered operator were found to cancel
in the ratio $\ratio{a}$, because supersymmetry enforces
equal rescalings of the gaugino mass and gauge coupling squared.
Finally, corrections unique to supergravity were discussed,
where they were shown to depend on the form of the gauge kinetic 
function and the K\"{a}hler potential.

Generating gaugino masses by gauge-mediation was discussed in the
final section.  We calculated the one-loop gaugino masses 
induced by a heavy chiral multiplet, in agreement with other 
well-known calculations when the mass is induced entirely from
gauge-mediation.  Since the existence of a messenger sector and the
size of the supersymmetry breaking masses is unknown, there
can be arbitrarily large ``corrections'' to the ratio $\ratio{a}$.
If gauge-mediated supersymmetry breaking at low energy is the only 
source of the gaugino mass, then the two-loop effects considered 
in this paper are expected to be much smaller than a supergravity
model, because the RG evolution is considerably reduced.

We should point out that many of the effects that we have discussed
could occur simultaneously (either constructively 
or destructively).  Determining the origin of the corrections to 
$\ratio{a}$ would therefore be quite difficult.  However, with 
sufficiently precise measurements of the weak scale spectrum it is 
expected that some of the possible disruptive effects could be 
ruled out.  Extracting the values of $\ratio{a}$ unambiguously at
the high scale relies on finding techniques that can set limits or 
exclude the myriad effects discussed in this paper.  
Since these effects include corrections from e.g.\ extensions of the MSSM 
and high scale effects, an alternative point of view is that 
the ratios $\ratio{a}$ are potentially sensitive to such physics.

Since the supersymmetric parameters $(g, M, Y, A)$ 
are sufficient to compute the RG evolution of $\ratio{a}$,
there was no need to specify any other parameters such as
squark or slepton soft masses.  This allowed us to compute various
corrections without any significant model dependence, 
although particular models were used to illustrate the
effects of an extra $U(1)$.  Since no attempt was made to 
construct complete extended models, the corrections to the 
MSSM ratios $\ratio{a}$ should be considered as generic expectations.
Some particular weak scale scenarios (or particular sets of 
parameters) that we have presented may be untenable when embedded 
in a more complete model for reasons such as 
vacuum stability, electroweak symmetry breaking, constraints from 
colliders, and constraints from cosmology.
Nevertheless, unambiguous extrapolation to the high scale can be 
expected to be difficult regardless of the model framework, 
given the varied effects that impact the one-loop invariant 
$\ratio{a}$.  Careful consideration of the effects discussed
in this paper would be necessary to make an unambiguous connection
between weak scale and high scale physics.

\section*{Acknowledgments}
\indent

I am very grateful to G.L.~Kane for many discussions and 
encouragement that led to this work, and also to S.P.~Martin
and J.D.~Wells for useful comments on the manuscript.
I also thank M.~Brhlik, H.-C.~Cheng, L.~Everett, C.~Kolda, 
S.P.~Martin, H.~Murayama, and E.~Poppitz for useful correspondence.
This work was supported in part by a Rackham predoctoral fellowship
and the U.S. Department of Energy.

\begin{appendix}
\refstepcounter{section}

\section*{Appendix~\thesection:~~Relevant $\beta$-functions}
\label{RGE-app}

\subsection{Gauge and gaugino $\beta$-functions}
\indent

The $\beta$-functions for the gauge couplings, gaugino masses,
and the ratio $M_a/\alpha_a$ in the $\overline{\mathrm{DR}}$ 
scheme~\cite{DRED} have been given before in 
Refs.~\cite{YamadaGaugino, MVGaugino, YamadaRatio, MVFull, 
YamadaFull}
and we generally follow the notation of Ref.~\cite{MVFull}:
\begin{eqnarray}
\frac{d}{dt} g_a &=& \frac{g_a^3}{16 \pi^2} B_a^{(1)} +
    \frac{g_a^3}{(16 \pi^2)^2} \left[ \sum_b B_{ab}^{(2)} g_b^2 
    - \sum_x C_a^x \tr (Y_x^\dagger Y_x) \right] \\
\frac{d}{dt} M_a &=& \frac{2 g_a^2}{16 \pi^2} B_a^{(1)} M_a +
    \frac{2 g_a^2}{(16 \pi^2)^2} \left[ \sum_b B_{ab}^{(2)} g_b^2 (M_a + M_b) 
    + \sum_x C_a^x (A_x - M_a) \tr (Y_x^\dagger Y_x) \right] \; .
\end{eqnarray}
Consequently, the two-loop RG equation for the ratio $M_a/\alpha_a$ is
\begin{eqnarray}
\frac{d}{dt} \frac{M_a}{g_a^2} &=& 
    \frac{2}{(16 \pi^2)^2} \left[ \sum_b B_{ab}^{(2)} g_b^2 M_b
    + \sum_x C_a^x A_x \tr (Y_x^\dagger Y_x) \right]
\label{ratio-RGE-eq}
\end{eqnarray}
where $1 \le a, b \le 3$ for the MSSM [$1 \le a,b \le 4$ for the MSSM 
plus an additional $U(1)$; the $a=4$ elements of $g_a$ and $M_a$ 
correspond to the gauge coupling $g'$ and gaugino mass $M'$ 
associated with the $U(1)'$],
and $x = u, d, e (, S, D)$ for the Yukawa couplings 
$\mathbf{Y}_u, \mathbf{Y}_d, \mathbf{Y}_e (, Y_S, Y_D)$
and scalar trilinear couplings
$\mathbf{A}_u, \mathbf{A}_d, \mathbf{A}_e (, A_S, A_D)$.
(The notation $A_x \tr (Y_x^\dagger Y_x)$ is understood to mean
$A_x^{ij} {Y_x^{ji}}^\dagger Y_x^{ij}$ for the flavor indices $i, j$.)
At least the first three Yukawa couplings and trilinear couplings
$(u, d, e)$ are matrices in family space, however we only
consider third generation couplings to be nonzero.
The one-loop and two-loop coefficients are written as
\begin{eqnarray*}
B_a^{(1)}    &=& (b_1, b_2, b_3, b_4) \\
B_{ab}^{(2)} &=& \left( \begin{array}{cccc} 
    b_{11} & b_{12} & b_{13} & b_{14} \\
    b_{21} & b_{22} & b_{23} & b_{24} \\
    b_{31} & b_{32} & b_{33} & b_{34} \\
    b_{41} & b_{42} & b_{43} & b_{44}
    \end{array} \right) \\
C_a^{u,d,e,S,D} &=& \left( \begin{array}{ccccc} 
    \textfrac{26}{5} & \textfrac{14}{5} & \textfrac{18}{5} &
        \textfrac{6}{5} & \textfrac{4}{5} \\
    6 & 6 & 2 & 2 & 0 \\
    4 & 4 & 0 & 0 & 2 \\
    c_{4u} & c_{4d} & c_{4e} & c_{4S} & c_{4D}
    \end{array} \right) \; .
\end{eqnarray*}
With additional matter may also come additional Yukawa couplings
in the superpotential, and thus $C_a^x$ is potentially an
$3 \times m$ matrix for the MSSM [$4 \times m$ matrix for the MSSM 
plus an additional $U(1)$] with $m$ distinct Yukawa couplings.
The coefficients $b_a$, $b_{ab}$ and $c_{4x}$
\begin{eqnarray*}
b_1 &=& \textfrac{1}{10} \left( n_Q + 8 n_u + 2 n_d + 3 n_L + 6 n_e 
    + 3 n_1 + 3 n_2 \right) \\
b_2 &=& \textfrac{1}{2} \left( 3 n_Q + n_L + n_1 + n_2 \right) 
    + 2 n_3 - 6 \\
b_3 &=& \textfrac{1}{2} \left(2 n_Q + n_u + n_d \right) 
    + 3 n_8 - 9 \\
b_4 &=& 6 n_Q Q_Q^2 + 3 n_u Q_u^2 + 3 n_d Q_d^2 + 2 n_L Q_L^2 + n_e Q_e^2
    + 2 n_1 Q_1^2 + 2 n_2 Q_2^2 + n_S Q_S^2 \\
b_{11} &=& \textfrac{1}{150} \left( n_Q + 128 n_u + 8 n_d + 27 n_L 
    + 216 n_e + 27 n_1 + 27 n_2 \right) \\
b_{12} &=& \textfrac{3}{10} \left( n_Q + 3 n_L + 3 n_1 + 3 n_2 \right) \\
b_{13} &=& \textfrac{8}{15} \left( n_Q + 8 n_u + 2 n_d \right) \\
b_{14} &=& \textfrac{2}{5} \left( n_Q Q_Q^2 + 8 n_u Q_u^2 + 2 n_d Q_d^2
    + 3 n_L Q_L^2 + 6 n_e Q_e^2 + 3 n_1 Q_1^2 + 3 n_2 Q_2^2 \right) \\
b_{21} &=& \textfrac{1}{10} \left( n_Q + 3 n_L + 3 n_1 + 3 n_2 \right) \\
b_{22} &=& \textfrac{7}{2} \left( 3 n_Q + n_L + n_1 + n_2 \right)
    + 24 n_3 - 24 \\
b_{23} &=& 8 n_Q \\
b_{24} &=& 2 \left( 3 n_Q Q_Q^2 + n_L Q_L^2 + n_1 Q_1^2 + n_2 Q_2^2 \right) \\
b_{31} &=& \textfrac{1}{15} \left( n_Q + 8 n_u + 2 n_d \right) \\
b_{32} &=& 3 n_Q \\
b_{33} &=& \textfrac{17}{3} \left( 2 n_Q + n_u + n_d \right) 
    + 42 n_8 - 54 \\
b_{34} &=& 2 \left( 2 n_Q Q_Q^2 + n_u Q_u^2 + n_d Q_d^2 \right) \\
b_{41} &=& \textfrac{2}{5} \left( n_Q Q_Q^2 + 8 n_u Q_u^2 + 2 n_d Q_d^2
    + 3 n_L Q_L^2 + 6 n_e Q_e^2 + 3 n_1 Q_1^2 + 3 n_2 Q_2^2 \right) \\
b_{42} &=& 3 \left( 6 n_Q Q_Q^2 + 2 n_L Q_L^2 + 2 n_1 Q_1^2 + 2 n_2 Q_2^2 
    \right) \\
b_{43} &=& 16 \left( 2 n_Q Q_Q^2 + n_u Q_u^2 + n_d Q_d^2 
    \right) \\
b_{44} &=& 4 \left( 6 n_Q Q_Q^4 + 3 n_u Q_u^4 + 3 n_d Q_d^4 + 2 n_L Q_L^4
    + n_e Q_e^4 + 2 n_1 Q_1^4 + 2 n_2 Q_2^4 + n_S Q_S^4 
    \right) \\
c_{4u} &=& 12 \left( Q_Q^2 + Q_u^2 + Q_2^2 \right) \\
c_{4d} &=& 12 \left( Q_Q^2 + Q_d^2 + Q_1^2 \right) \\
c_{4e} &=& 4 \left( Q_L^2 + Q_e^2 + Q_1^2 \right) \\
c_{4s} &=& 4 \left( Q_S^2 + Q_1^2 + Q_2^2 \right) \\
c_{4D} &=& 6 \left( Q_S^2 + Q_D^2 + Q_{D^c}^2 \right)
\end{eqnarray*}
in terms of the MSSM multiplets $(Q, u^c, d^c, L, e^c, H_1, H_2)$, 
an MSSM gauge singlet $(S)$, and the (exotic) 
adjoints $\mathbf{3}$ and $\mathbf{8}$.
The charge assignments to $SU(3)_c$, $SU(2)_L$, $U(1)_Y$,
and $U(1)'$ and multiplicities $n_i$ of the matter supermultiplets
are shown in Table~\ref{charge-table}.
To incorporate the extra matter that exists in a $\mathbf{27}$ 
of $E_6$, let $n_d Q_d^n \ra n_d Q_d^n + n_D Q_D^n + n_{D^c} Q_{D^c}^n$
and $n_S Q_S^n \ra n_S Q_S^n + n_{\nu^c} Q_{\nu^c}^n$ in the
above equations for $n=0, 2, 4$.  
Note that for the MSSM without a singlet, we could 
have ignored the distinction between $n_1$ and $n_2$, and written
the above in terms of just $n_H = n_1 = n_2$.  However, the
presence of the term $Y_S S H_1 H_2$ as a substitute for $\mu H_1 H_2$
in the superpotential implies $Q_1 \ne - Q_2$, 
and thus $n_1$ and $n_2$ must be treated separately.
\begin{table}
\renewcommand{\baselinestretch}{1.2}\small\normalsize
\begin{center}
\begin{tabular}{r|ccccccccccc}
    & $Q$ & $u^c$ & $d^c$ & $L$ & $e^c$ & $H_1$ & $H_2$ 
    & $D$ & $D^c$ & $\nu^c$ & $S$ \\ \hline
multiplicity & $n_Q$ & $n_u$ & $n_d$ & $n_L$ & $n_e$ & $n_1$ & $n_2$ 
    & $n_D$ & $n_{D^c}$ & $n_{\nu^c}$ & $n_S$ \\ \hline
$SU(3)_c$ & $\mathbf{3}$ & $\mathbf{\overline{3}}$ & $\mathbf{\overline{3}}$ &
    $\mathbf{1}$ & $\mathbf{1}$ & $\mathbf{1}$ & $\mathbf{1}$ &
    $\mathbf{3}$ & $\mathbf{\overline{3}}$ & $\mathbf{1}$ & $\mathbf{1}$ \\
$SU(2)_L$ & $\mathbf{2}$ & $\mathbf{1}$ & $\mathbf{1}$ & $\mathbf{2}$ &
    $\mathbf{1}$ & $\mathbf{2}$ & $\mathbf{2}$ & $\mathbf{1}$ &
    $\mathbf{1}$ & $\mathbf{1}$ & $\mathbf{1}$ \\
$Y/2$ of $U(1)_Y$ & 
    $\textfrac{1}{6}$ & $-\textfrac{2}{3}$ &
    $\textfrac{1}{3}$ & $-\textfrac{1}{2}$ & $1$ & $-\textfrac{1}{2}$ &
    $\textfrac{1}{2}$ & $-\textfrac{1}{3}$ & $\textfrac{1}{3}$ & $0$ & $0$ \\
$Q$ of $U(1)'$ & $Q_Q$ & $Q_u$ & $Q_d$ & $Q_L$ & $Q_e$ & 
    $Q_1$ & $Q_2$ & $Q_D$ & $Q_{D^c}$ & $Q_{\nu^c}$ & $Q_S$ \\
$2 \sqrt{15} Q_\eta$ of $U(1)_\eta$ & $-2$ & $-2$ & $1$ & $1$ & $-2$ & 
    $1$ & $4$ & $4$ & $1$ & $-5$ & $-5$ \\ \hline
\end{tabular}
\end{center}
\renewcommand{\baselinestretch}{1.0}\small\normalsize
\caption{Charge assignments for the matter considered in this
paper.  We did not list the $\mathbf{3}$ and $\mathbf{8}$ with 
multiplicities $n_3$ and $n_8$ that are in the adjoints of 
$SU(2)_L$ and $SU(3)_c$ respectively, since they are singlets under 
all other groups.  The $\eta$-model of $E_6$ is a special 
case of a general $U(1)'$ extension with the charges shown.  
Note that
$Q_{\mathrm{em}} = T_3 + Y/2$, and $U(1)_Y$ is in the GUT 
normalization.}
\label{charge-table}
\end{table}

We illustrate the utility of the above expressions by calculating
the $\beta$-function coefficients $B^{(1)}$ and $B^{(2)}$
for a few particular cases.  First, for the MSSM
\begin{eqnarray*}
n_Q &=& n_u \; = \; n_d \; = \; n_L \; = \; n_e \; = \; 3 \\
n_1 &=& n_2 \; = \; 1 \\
n_3 &=& n_8 \; = \; n_S \; = \; 0 
\end{eqnarray*}
we obtain the very well-known result
\begin{eqnarray}
B_a^{(1)}    &=& \left( \textfrac{33}{5}, 1, -3 \right) \\
B_{ab}^{(2)} &=& \left( \begin{array}{ccc} 
    \textfrac{199}{25} & \textfrac{27}{5} & \textfrac{88}{5} \\
    \textfrac{9}{5}    & 25               & 24 \\
    \textfrac{11}{5}   & 9                & 14 \\
    \end{array} \right) \; .
\end{eqnarray}
For a model with additional $\mathbf{5}$'s or $\mathbf{\overline{5}}$'s
of SU(5) that transform as $(\mathbf{5}, Q_{\mathbf{5}})$ and
$(\mathbf{\overline{5}}, -Q_{\mathbf{5}})$ under [$SU(5)$, $U(1)'$], 
one can compute the shift in the one-loop and two-loop
$\beta$-functions
\begin{eqnarray}
\Delta B_a^{(1)} &=& \left( \textfrac{1}{2}, \textfrac{1}{2}, 
    \textfrac{1}{2}, 5 Q_{\mathbf{5}}^2 \right) n_{\mathbf{5}} \\
\Delta B_{ab}^{(2)} &=& \left( \begin{array}{cccc}
    \textfrac{7}{30} & \textfrac{9}{10} & \textfrac{16}{15} 
        & 2 Q_{\mathbf{5}}^2 \\
    \textfrac{3}{10} & \textfrac{7}{2} & 0 & 2 Q_{\mathbf{5}}^2 \\
    \textfrac{2}{15} & 0 & \textfrac{17}{3} & 2 Q_{\mathbf{5}}^2 \\
    2 Q_{\mathbf{5}}^2 & 6 Q_{\mathbf{5}}^2 & 16 Q_{\mathbf{5}}^2 
        & 20 Q_{\mathbf{5}}^4 \\
    \end{array} \right) n_{\mathbf{5}}
\end{eqnarray}
and similarly for additional $\mathbf{10}$'s or $\mathbf{\overline{10}}$'s
of SU(5) that transform as $(\mathbf{10}, Q_{\mathbf{10}})$ and
$(\mathbf{\overline{10}}, -Q_{\mathbf{10}})$ under [$SU(5)$, $U(1)'$], 
\begin{eqnarray}
\Delta B_a^{(1)} &=& \left( \textfrac{3}{2}, \textfrac{3}{2}, 
    \textfrac{3}{2}, 10 Q_{\mathbf{10}}^2 \right) n_{\mathbf{10}} \\
\Delta B_{ab}^{(2)} &=& \left( \begin{array}{cccc}
    \textfrac{23}{10} & \textfrac{3}{10} & \textfrac{24}{5} 
        & 6 Q_{\mathbf{10}}^2 \\
    \textfrac{1}{10} & \textfrac{21}{2} & 8 & 6 Q_{\mathbf{10}}^2 \\
    \textfrac{3}{5} & 3 & 17 & 6 Q_{\mathbf{10}}^2 \\
    6 Q_{\mathbf{10}}^2 & 18 Q_{\mathbf{10}}^2 & 48 Q_{\mathbf{10}}^2 
        & 40 Q_{\mathbf{10}}^4 \\
    \end{array} \right) n_{\mathbf{10}}
\end{eqnarray}
Note that we chose the simplest assignment of the $U(1)'$ charge,
such that it is assigned vectorially and commutes with $SU(5)$.
Other assignments can be handled by simply taking the appropriate 
components of the decomposed SU(5) field individually, and
assigning U(1)$'$ charge as desired (for examples of other $U(1)'$ 
charge assignments, see Ref.~\cite{BKMR}).  The 
[$SU(3)_c$, $SU(2)_L$, $U(1)_Y$] pieces of the above 
$\beta$-functions were also calculated in e.g.\ 
Refs.~\cite{CaroneMurayama, KoldaMR, GLR}, and we agree 
with their results.

Finally, we use the above results to compute the $\beta$-function 
coefficients for the $\eta$-model of $E_6$ with three $\mathbf{27}$ 
matter representations.  The charge assignments of the $U(1)_\eta$
are shown in Table~\ref{charge-table}.  We obtain
\begin{eqnarray}
B_a^{(1)}    &=& \left( 9, 3, 0, 9 \right) \\
B_{ab}^{(2)} &=& \left( \begin{array}{cccc} 
    9 &  9 & 24 & 3 \\
    3 & 39 & 24 & 3 \\
    3 &  9 & 48 & 3 \\
    3 &  9 & 24 & 9 \\
    \end{array} \right) \; .
\end{eqnarray}
The one-loop and two-loop $\beta$-function coefficients were 
also given in Ref.~\cite{EENZ, DVJ}.  We agree completely with the
results in Ref.~\cite{DVJ}, and we agree with the results 
in Ref.~\cite{EENZ} except for the overall sign and the values 
of the two-loop elements $B^{(2)}_{23}$, $B^{(2)}_{14}$, 
and $B^{(2)}_{41}$ (in our notation).

\subsection{Thresholds}
\label{thresholds-app}
\indent

In running the RG equations, we will encounter two kinds of
thresholds.  The first, what we call ``fully supersymmetric
thresholds'', occur when extra matter is decoupled from
the spectrum in a fully supersymmetry way.  The RG equations
as given above are sufficient to remove matter (both the
scalar and fermionic components of a chiral superfield) 
at any scale, provided the masses of the scalar and 
fermionic components are the same.  Both one-loop and two-loop
decoupling can be achieved; the importance of the latter
for the $M_a/\alpha_a$ ratio is discussed in Sec.~\ref{extra-matter-sec}.  
The second kind of thresholds are those near the weak scale,
when the supersymmetric components of the SM spectrum
are decoupled.  One way to treat weak scale thresholds is
to decouple the supersymmetric components from the 
$\beta$-functions, creating non-supersymmetric effective
theories.  Expressions for the corrections to the 
one-loop $\beta$-functions are given in Ref.~\cite{CPR}, 
and we have translated their results into the formalism 
we have used above:
\begin{eqnarray*}
n_Q &\ra&   \textfrac{1}{3} n_u^{(f)} + \textfrac{1}{3} n_d^{(f)}
          + \textfrac{1}{6} n_{\tilde{u}_L}^{(s)} 
          + \textfrac{1}{6} n_{\tilde{d}_L}^{(s)} \\
n_u &\ra&   \textfrac{2}{3} n_u^{(f)} 
          + \textfrac{1}{3} n_{\tilde{u}_R}^{(s)} \\
n_d &\ra&   \textfrac{2}{3} n_d^{(f)} 
          + \textfrac{1}{3} n_{\tilde{d}_R}^{(s)} \\
n_L &\ra&   \textfrac{1}{3} n_v^{(f)} + \textfrac{1}{3} n_e^{(f)}
          + \textfrac{1}{6} n_{\tilde{v}}^{(s)} 
          + \textfrac{1}{6} n_{\tilde{e}_L}^{(s)} \\
n_e &\ra&   \textfrac{2}{3} n_e^{(f)} 
          + \textfrac{1}{3} n_{\tilde{e}_R}^{(s)} \\
n_1 &\ra&   \textfrac{2}{3} n_{\tilde{H}_1}^{(f)}
          + \textfrac{1}{3} n_{H_1}^{(s)} \\
n_2 &\ra&   \textfrac{2}{3} n_{\tilde{H}_2}^{(f)}
          + \textfrac{1}{3} n_{H_2}^{(s)}
\end{eqnarray*}
where $f$ is the fermion and $s$ is the complex scalar.
In addition, the $\beta$-function coefficients for the non-Abelian
gauge couplings have the further changes
\begin{eqnarray*}
b_2: \quad -6 &\ra& - \textfrac{22}{3} + \textfrac{4}{3} \theta_{\tilde{W}} \\
b_3: \quad -9 &\ra& - 11 + 2 \theta_{\tilde{g}}
\end{eqnarray*}
where $\theta_{\tilde{W}, \tilde{g}} = 0$ or $1$ if the renormalization
scale $Q$ is less than or greater than $|M_{\tilde{W}, \tilde{g}}|$
respectively.  Finally, note that the two-loop corrections due 
to weak scale thresholds could in principle be calculated, 
but we ignore these corrections since they are not enhanced
by a large logarithm (unlike fully supersymmetric thresholds),
see Ref.~\cite{Hall} for further discussion.

\subsection{Yukawa and scalar trilinear $\beta$-functions}
\label{Yukawa-trilinear-app}
\indent

The general form of the one-loop RG equations for the 
Yukawa couplings and soft scalar trilinear couplings is
\begin{eqnarray}
\frac{d}{dt} Y_i &=& \frac{Y_i}{16 \pi^2} \left[ \sum_j c_{ij} Y_j^2 
    - \sum_k d_{ik} g_k^2 \right] \\
\frac{d}{dt} A_i &=& \frac{1}{8 \pi^2} \left[ \sum_j c_{ij} Y_j^2 A_j 
    + \sum_k d_{ik} g_k^2 M_k \right]
\end{eqnarray}
where the coefficients are
\begin{eqnarray*}
c_{ij} &=& \left( \begin{array}{ccccc} 
            6 & 1 & 0 & 1 & 0 \\
            1 & 6 & 1 & 1 & 0 \\
            0 & 3 & 4 & 1 & 0 \\
            3 & 3 & 1 & 4 & 3 \\
            0 & 0 & 0 & 2 & 5 \\
            \end{array} \right) \\
d_{ik} &=& \left( \begin{array}{cccc}
    \textfrac{13}{15} & 3 & \textfrac{16}{3} & 2 (Q_Q^2 + Q_u^2 + Q_2^2) \\
    \textfrac{7}{15}  & 3 & \textfrac{16}{3} & 2 (Q_Q^2 + Q_d^2 + Q_1^2) \\
    \textfrac{9}{5}   & 3 & 0                & 2 (Q_L^2 + Q_e^2 + Q_1^2) \\
    \textfrac{3}{5}   & 3 & 0                & 2 (Q_S^2 + Q_1^2 + Q_2^2) \\
    \textfrac{2}{15}  & 0 & \textfrac{16}{3} & 2 (Q_S^2 + Q_D^2 + Q_{D^c}^2) \\
    \end{array} \right) 
\end{eqnarray*}
for $i,j$ = $(u,d,e,S,D)$ and $k = 1,2,3,4$.  We are neglecting
the first and second generation Yukawa couplings and associated
mixings between generations.
Two-loop corrections to the Yukawa couplings~\cite{BBO1, CPR, 
MVFull, YamadaFull} 
and the associated scalar trilinear couplings~\cite{MVFull, YamadaFull} 
have been calculated, and expressions for the two-loop corrections 
to $Y_S$ and $Y_D$ could be easily calculated from 
Refs.~\cite{MVFull, YamadaFull}.  We consistently neglect
two-loop corrections to Yukawa couplings since they are of 
the same order as three-loop corrections (which we also neglect) 
to the main observables $g_a$, $M_a$, and $M_a/\alpha_a$ 
relevant to this paper.

\refstepcounter{section}
\section*{Appendix~\thesection:~~Cancellation of $U(1)'$ anomalies}
\label{anomalies-app}

\subsection{General conditions}
\indent

There are six anomalies associated with enlarging the MSSM
gauge group to include an extra $U(1)'$ that need to be 
canceled if we expect a sensible low energy
effective theory.  These are the mixed gauge anomalies,
$[SU(3)_c]^2 U(1)'$, $[SU(2)_L]^2 U(1)'$, $[U(1)_Y]^2 U(1)'$, 
$[U(1)']^2 U(1)_Y$, the triangle anomaly $[U(1)']^3$, and
the mixed gravitational anomaly $U(1)' [\mathrm{gravity}]^2$.  With
the addition of the mixed gravitational anomaly, this appendix
follows closely the discussion of $U(1)'$ anomalies in 
Ref.~\cite{Langacker}.

With the particle content, multiplicities, and $U(1)'$ charges 
stated in Table~\ref{charge-table}, the cancellation of the 
$[SU(3)_c]^2 U(1)'$, $[SU(2)_L]^2 U(1)'$, $[U(1)_Y]^2 U(1)'$,
$[U(1)']^2 U(1)_Y$, $[U(1)']^3$, $U(1)' [\mathrm{gravity}]^2$
anomalies require, respectively,
\begin{eqnarray*}
0 &=& 2 n_Q Q_Q + n_u Q_u + n_d Q_d + n_D Q_D + n_{D^c} Q_{D^c} \\
0 &=& 3 n_Q Q_Q + n_L Q_L + n_1 Q_1 + n_2 Q_2 \\
0 &=& 2 n_Q Q_Q + 16 n_u Q_u + 4 n_d Q_d + 6 n_L Q_L + 12 n_e Q_e
    + 6 n_1 Q_1 + 6 n_2 Q_2 + 4 n_D Q_D + 4 n_{D^c} Q_{D^c} \\
0 &=& n_Q Q_Q^2 - 2 n_u Q_u^2 + n_d Q_d^2 - n_L Q_L^2 + n_e Q_e^2
    - n_1 Q_1^2 + n_2 Q_2^2 - n_D Q_D^2 + n_{D^c} Q_{D^c}^2 \\
0 &=& 6 n_Q Q_Q^3 + 3 n_u Q_u^3 + 3 n_d Q_d^3 + 2 n_L Q_L^3 + n_e Q_e^3
    + 2 n_1 Q_1^3 + 2 n_2 Q_2^3 + 3 n_D Q_D^3 + 3 n_{D^c} Q_{D^c}^3 \\
& &{} + n_S Q_S^3 + n_{\nu^c} Q_{\nu^c}^3 \\
0 &=& 6 n_Q Q_Q + 3 n_u Q_u + 3 n_d Q_d + 2 n_L Q_L + n_e Q_e
    + 2 n_1 Q_1 + 2 n_2 Q_2 + 3 n_D Q_D + 3 n_{D^c} Q_{D^c} \\
& &{} + n_S Q_S + n_{\nu^c} Q_{\nu^c} \; .
\end{eqnarray*}
For completeness, the further constraint of preventing radiatively 
generated kinetic mixing $\tr(Y Q) = 0$ is
\begin{eqnarray*}
0 &=& n_Q Q_Q - 2 n_u Q_u + n_d Q_d - n_L Q_L + n_e Q_e
    - n_1 Q_1 + n_2 Q_2 - n_D Q_D + n_{D^c} Q_{D^c} \; .
\end{eqnarray*}

\subsection{Minimal $U(1)'$ model}
\label{minimal-app}
\indent

The model of Ref.~\cite{Langacker} has the superpotential
\begin{eqnarray}
W &=& Y_t Q H_2 u^c + Y_S S H_1 H_2 \; ,
\end{eqnarray}
(devoid of Yukawa couplings to the $b$ or $\tau$ sector),
with the MSSM spectrum and one singlet $S$, and the MSSM 
group plus an additional $U(1)'$.
The authors of Ref.~\cite{Langacker} wrote down
one simple example of anomaly cancellation in their model,
\begin{eqnarray*}
Q_Q &=& -\textfrac{1}{3} Q_1 \\
Q_u &=& \textfrac{1}{3} (Q_1 - 3 Q_2) \\
Q_d &=& \textfrac{1}{3} (Q_1 + 3 Q_2) \\
Q_L &=& - Q_2 \\
Q_e &=& Q_2 - Q_1 \\
Q_S &=& -(Q_1 + Q_2) \; ,
\end{eqnarray*}
where the charges given above for $(Q, u, d, L, e)$ are assigned 
only to the third generation (first and second generations have
zero $U(1)'$ charge).  To prevent radiatively generated kinetic 
mixing, we also impose the constraint
\begin{eqnarray}
Q_2 &=& \textfrac{4}{9} Q_1 \; .
\label{Q1Q2constraint-eq}
\end{eqnarray}
Thus all of the charges are determined up to one ``overall'' charge,
which we take to be $Q_1$.  Since the gauge coupling $g'$ is
always accompanied one power of some $U(1)'$ charge, we
can therefore choose to rescale either $Q_1$ or $g'$, without 
loss of generality.  If the $U(1)'$ were embedded in a higher
rank group at the high scale, then the rescaling factor would 
be fixed by the particular embedding.

Note that experimental bounds on $Z$--$Z'$ mixing also impose 
constraints on the size of $g'Q_1$, and may not even allow 
Eq.~(\ref{Q1Q2constraint-eq}) to be satisfied in the particular
scenarios where the vev of $S$ is comparable to the vevs $v_1$
and $v_2$ (from $H_1$ and $H_2$)~\cite{Langacker}.  
The mixing angle in the $Z$--$Z'$ system is given by
\begin{eqnarray}
\tan 2 \alpha_{Z - Z'} &=& \frac{g' \sqrt{g_1^2 + g_2^2} 
   (v_1^2 Q_1 - v_2^2 Q_2)}{M_{Z'}^2 - M_Z^2} \; .
\end{eqnarray}
We find the intriguing result that simultaneously satisfying
Eq.~(\ref{Q1Q2constraint-eq}), and zero mixing 
occurs when $\tan\beta = \textfrac{3}{2}$.  Of course the mixing 
could be nonzero but simply suppressed by a large $Z'$ mass, although
generally $Q_1 Q_2 > 0$ is needed for a partial 
cancellation in the numerator of the above expression [which
is compatible with Eq.~(\ref{Q1Q2constraint-eq})].
In any case, for our purposes we will take
Eq.~(\ref{Q1Q2constraint-eq}) without specifying 
$\tan\beta$, and assume that appropriate (larger) vevs of $S$ 
are induced as necessary to avoid the experimental bounds.

\subsection{$E_6$ model}
\label{E6-app}
\indent

Using the charge assignments given in Table~\ref{charge-table},
it is easy to show that the $\eta$-model of $E_6$ with three 
generations of $\mathbf{27}$s cancels all of the anomalies above.  
This is not at all surprising, since it is well known that 
theories with matter in complete representations of a non-Abelian 
group are nonanomalous.  Three generations of $\mathbf{27}$s are 
sufficient to incorporate the matter content of the MSSM, but
also include two extra generations of up-type and down-type Higgs 
doublets, three generations of right-handed neutrinos, and
three generations of MSSM gauge singlets.

\end{appendix}


\begin{thebibliography}{99}
\bibitem{HMGnote}
    The sign in front of the $1/4\pi$ is consistent with Eq.~(10),
    but opposite to that in Eq.~(15) of Ref.~\cite{HMG}.
    The sign is consistent with an equivalent expression 
    given in Ref.~\cite{KoldaMR}.
\end{thebibliography}
\end{document}